\documentclass[%
reprint,superscriptaddress,
amsmath,amssymb,
pre,
aps,  
showkeys,
nofootinbib% a enlever eventuellement
]{revtex4-2}

\usepackage{hyperref}
\usepackage{graphicx}
\usepackage{amsfonts} 
\usepackage{thmbox} %antoine
\usepackage{empheq} %nouveau
\usepackage{version}
\usepackage{times}
\usepackage{subfigure}
\usepackage{textcomp}
\usepackage{makeidx}
\usepackage{enumerate}
\usepackage{amsmath}
\usepackage{fancyhdr}
\usepackage[top=20mm, bottom=30mm, left=15mm, right=15mm]{geometry}
\usepackage{color}

\newcommand{\beq}{\begin{equation}}     \newcommand{\eeq}{\end{equation}}
\newcommand{\beqa}{\begin{eqnarray}}    \newcommand{\eeqa}{\end{eqnarray}}
\newcommand{\bde}{\begin{description}}  \newcommand{\ede}{\end{description}}
\newcommand{\ben}{\begin{enumerate}}    \newcommand{\een}{\end{enumerate}}

\newcommand{\clp}{\clearpage }

\newcommand{\kT}{{k_{\rm B}T} } 
\newcommand{\kB}{{k_{\rm B}} } 
\newcommand{\gtsim}{\mbox{${}^{>}_\sim$}}

\newcommand{\R}{{\mathbb{R}}} 
 
\newcommand{\Z}{{\mathbb{Z}}} 
\newcommand{\eqn}[1]{\beq{ #1 }\eeq}

\newcommand{\inv}[1]{{\frac{1}{#1}}}

\newcommand{\inRbracket}[1]{{\left({#1}\right)}}

\newcommand{\sech}{{\mbox{sech}}}

\newtheorem[L]{thm}{Theorem}[section]
\newtheorem{cor}[thm]{Corollary}
\newtheorem{theorem}{{\sf Assertion :}}[section] 
\newtheorem{definition}{\sf Definition} 
 
\newtheorem{lemma}[theorem]{Lemma}
\newcommand{\bth}{\begin{thm}}  
\newcommand{\blem}{\begin{lemma}}  
	\newcommand{\elem}{\end{lemma}}  
\newcommand{\bpr}{\begin{proof}}  
	\newcommand{\epr}{\end{proof}} 
\newcommand{\bdefine}{\begin{definition}}  
	\newcommand{\edefine}{\end{definition}}  
\newcommand{\bcor}{\begin{cor}} 
	\newcommand{\ecor}{\end{cor}}  
\newcommand{\bprop}{\begin{example}[Property]}  
	\newcommand{\eprop}{\end{example}}  
\newcounter{formulaire}
\newcommand{\beqf}{\addtocounter{formulaire}{1}\begin{equation}}
	\newcommand{\eeqf}{\tag{R \arabic{formulaire}}\end{equation}}
\newcommand{\beqaf}{\addtocounter{formulaire}{1}\begin{equation}\begin{array}{rcl}}
		\newcommand{\eeqaf}{\end{array}\tag{R \arabic{formulaire}}\end{equation}}

\newcommand{\meq}{{m^\mathrm{eq}}}

\newcommand{\pair}[2]{{\begin{matrix} {#1} \\  {#2} \\   \end{matrix}}}

\usepackage{dcolumn}
\usepackage{bm}
\usepackage{listings}
\usepackage{ulem}
\usepackage{graphicx}

\begin{document}
	
	\title{Interplay between Markovianity and Progressive Quenching}
	\author{Charles Moslonka}
	\affiliation{Laboratoire Gulliver, UMR CNRS 7083, ESPCI Paris, Universit\'e PSL\\
		10 rue Vauquelin, 75005, Paris, France.}
	\email[Corresponding author: ]{charles.moslonka@espci.psl.eu}
	\author{Ken Sekimoto}
	\affiliation{Laboratoire Gulliver, UMR CNRS 7083, ESPCI Paris, Universit\'e PSL\\
		10 rue Vauquelin, 75005, Paris, France.}
	\affiliation{Laboratoire Mati\`ere et Syst\`emes Complexes, UMR CNRS 7057, Universit\'e Paris Cit\'e,\\
		10 Rue Alice Domon et Léonie Duquet, 75013, Paris, France }
	\email[Corresponding author: ]{ken.sekimoto@espci.psl.eu}
	
\begin{abstract}
Progressive quenching (PQ) is a process in which we sequentially fix a system's degrees of freedom, which would otherwise evolve according to their stochastic dynamics. Previous studies have discovered what we refer to as the hidden martingale property in PQ. Here, we first attribute this martingale property to the canonicity of the two-layer ensemble comprising quenched and thermal ensembles and demonstrate that the Markovian property, coupled with the detailed balance (DB) of the evolution dynamics, underpins this canonicity. We then expand the PQ to the Markovian dynamics on the transition network where the DB is not required. Additionally, we examine the PQ of the systems that evolve through non-Markovian dynamics between consecutive quenching. When non-Markovian dynamics ensure a trajectory-wise DB, such as in an equilibrium spin system with a hidden part, the PQ can occasionally maintain the canonical structure of the overall statistical ensemble, but not always. Lastly, we analytically and numerically investigate the PQ of a non-Markovian spin system with delayed interaction and illustrate how the reduction of spin correlations due to the delay can be compensated by the PQ.
\end{abstract}
\maketitle

\section{Introduction}

Markov chains have been studied in depth for more than a century now, whose fields of application are broad and diverse \cite{norris1998markov, grimmett2020probability,  rao_esposito_2016noneq_thermo_chemical_reaction, allen2010introduction_stoch_process_biology}. This is especially the case of stochastic physics \cite{jacobs2010stochastic, van_kampen1992stochastic, LNP}. 	Some studies concern the effect of change of the parameters of the Markov chain, that is, the topology of the transition network and the rates associated with the jump on it \cite{hill1966diagram, schnakenberg1976network, Harunari_Dutta_Polettini_Roldan_PRX2022_visible_transition, PhysRevX.12.031025_Udo}. What we call Progressive Quenching (PQ) belongs to this category of studies, where the system that was initially in equilibrium is modified by progressively fixing a part of the system's degrees of freedom.

In our previous work \cite{PQ-KS-BV-pre2018, PQ-CM-KS-2020, PQ-CM-KS-2022}, the results of which are briefly summarized in Section \ref{sec:model} for the sake of self-containedness, we demonstrated that a martingale process emerged during the Progressive Quenching of a model of Ising spins. We tested this result both numerically and theoretically, using the inverse system size expansion as well as tower-rule-based arguments under specific hypotheses. This property enabled us to make predictions about the individual future trajectories of the process, in addition to inferring the anterior one. We realized that there is a canonical distribution underlying the ensemble of the final quenched configurations, and we explored different approaches to understand its origin via an explicit construction of path weight through the "local invariance," which is equivalent to the martingale property that emerged during the evolution process \cite{PQ-CM-KS-2022}.

So far, we have assumed that before each quenching operation, the unquenched part of the system is in thermal equilibrium under a given number of fixed spins,
and we have implicitly assumed continuous-time Markovian dynamics. However, we have not been conscious enough in distinguishing the dynamic aspect from the statistical aspect when defining the quenching process. In particular, we have yet to fully appreciate the importance of the Markovian property of the evolution dynamics.

In the present paper, we want to understand the conditions under which the canonical characteristics of the whole ensemble (both quenched and unquenched) are conserved along the PQ process and how important is the Markovian assumption on the stochastic evolution for the martingale and underlying canonical structure.
We will discuss the Markovian and non-Markovian cases separately and distinguish the dynamics with or without detailed balance.

The organization of the present paper is as follows:
Section \ref{sec:model} presents our model setup (\ref{subsec:IIA}) and previous results (\ref{subsec:IIB}), where we explain the protocol of PQ, the property that we call hidden martingale. 
In Section \ref{sec:PQ_in_markov_models} we first introduce the two-story ensemble and argue that the detailed balance and the Markovian dynamics are required for this ensemble to be canonical (\ref{subsec:two-story}). Then, we extend the PQ defined on any Markovian transition network (\ref{subsec:Markov_DB}), where the condition of detailed balance is also relaxed (\ref{subsec:MarkovNonDB}).
Then, Section \ref{sec:PQ_Non_Markov} deals with the case of non-Markovian systems. 
We argue that the canonical structure that supported the hidden martingale of PQ is generally broken even if the (trajectory-wise) detailed balance is initially assured (\ref{subsec:nM-DB}). Finally, we focus on the spin model with delayed interaction (\ref{sec:Choi_ Huberman}) and show how the operation of PQ interferes with the non-Markovian delay through the time interval
between the consecutive quench.
Finally, Section \ref{sec:conlustion} summarizes the results, and a comparison is made with the linear voter model.

\section{Models and previous results} \label{sec:model}	
\subsection{Model setup}\label{subsec:IIA}
Our previous studies focused on a globally-coupled Ising spins model, consisting of $N_0$ spins interacting pairwise on a complete graph with a coupling constant $j = \frac{J}{N_0}$. 
The Hamiltonian reads :
\begin{equation}
\mathcal{H} = -j \sum_{i<j}^{N_0} s_i s_j = -\frac{j}{2} M_{tot}^2 
+\frac{j}{2}N_0,
\end{equation}
where $M_{tot} = \sum_{i=1}^{N_0} s_i$. 
In this system, we attribute spin indices $i$ so that, after the $T$-th Progressive Quenching operation  $(1\leq T \leq N_0),$ 
the first $T$ spins $\{s_1,\ldots,s_T\}$ constitute the \textit{quenched} part 
and the remaining $N_0-T$ spins, $\{s_{T+1},\ldots, s_{N_0}\},$ the \textit{unquenched} or \textit{free} part. 
We denote by $M_T$ the quenched magnetization at the $T$-th stage: 
$$ M_T = \sum_{i=1}^T s_i .$$ 
The Hamiltonian for the remaining part then reads : 
\begin{equation}
\mathcal{H}_T = -j \sum_{T+1\leq i<j \leq N_0 } s_i s_j  
-( j M_T+h) \sum_{i=T+1}^{N_0} s_i,
\end{equation}
where $h$ is the external field.
We denote by $Z_{T, M_T} $ the partition function of the partially quenched system
characterized by $\mathcal{H}_T$ above. Hereafter, we set the inverse temperature $\beta = (\kT)^{-1}=1$ through an appropriate unit of energy.
The mean magnetization of the \textit{free} part at constrained thermal equilibrium, which we denote by $m^{(eq)}_{T,M_T},$ reads 
\begin{equation}
m^{(eq)}_{T,M_T} = %\frac{1}{Z_{T,M_T}}\frac
\left.{\partial \ln Z_{T,M_T}}/{\partial h}\right|_{h\to 0}.
\end{equation}
With such a setting, the operation of PQ is to move and
transform a spin from the \textit{free} part into the \textit{quenched} part in such a way that the value of that spin is maintained as it is.
Therefore, if the system has $T$ quenched spins, the subsequently quenched spin $s_{T+1}$ satisfies 
\begin{equation}
\mathbb{E}[s_{T+1} | M_T] = m^{(eq)}_{T,M_T}.
\end{equation}
When we regard $T$, the number of fixed spins, as a ``time'', 
we have two stochastic processes : the one being $M_T,$ the magnetization of the \textit{quenched} part of the system,  
and the second being $m^{(eq)}_{T,M_T},$  the equilibrated mean spin
of the \textit{free} part.

\subsection{Brief review of the previous results}\label{subsec:IIB}
\paragraph{Hidden martingale}:
\null{ For self-containedness, we first recall the definition of the discrete-time martingale process. We call the stochastic process $\{Y_T\}$ martingale associated with the stochastic process $\{X_T\}$ if (i) as for the past, $Y_T$ is determined when $(X_0,\ldots, X_T)$ are given, and (ii) as for the future, $\mathbb{E}[{\null} Y_T|X_0,\ldots, X_s]=Y_s$ for $T\ge s.$}

We showed previously \cite{PQ-KS-BV-pre2018, PQ-CM-KS-2020, PQ-CM-KS-2022} that, if we let the free spins reach thermal equilibrium after each quenching step, or equivalently quench the spin at $\pm 1,$ respectively, with the probability, $P(s_{T+1} = \pm 1) = \inv{2}{(1 \pm m^{(eq)}_{T, M_T})}$, then the evolution of $m^{(eq)}_{T, M_{T}}$ satisfies the martingale law \cite{martingales_review2022}:
\begin{equation}
\mathbb{E}[m^{(eq)}_{T,M_{T}}|M_s] = m^{(eq)}_{s,M_s}\quad (T\ge s).
\end{equation}
\null{ Apparently $m^{(eq)}_{T,M_T}$ plays a role of $Y_T,$ while as $\{X_0,\ldots,X_s\}$ we put only $M_s$ because $\{M_s\}$ is Markovian and $m^{(eq)}_{s,M_s}$ is independent of 
$\{M_0,\dots, M_{s-1}\}$ in the present problem.
The martingale property of $m^{(eq)}_{s,M_s}$ is said ``hidden'' because it is not of the main process, $M_T.$}
\paragraph{Consequences of the hidden martingale}: 
This particular property gives a large amount of information about the evolution of $M_T$ \cite{PQ-KS-BV-pre2018, PQ-CM-KS-2020}, first allowing an approximate prediction of the final distribution
of $M_{T=N_0}$ given the one at the early stage, even before the distribution bimodality appears.
Secondly, the hidden martingale implies an invariance of the path probabilities under a local trajectory modification. This property 
allowed us to derive a thermodynamic decomposition of the probability distribution of $M_{N_0}$, linking it to the canonical ensemble \cite{PQ-CM-KS-2022}. 
More concretely, the probability of observing the quenched magnetization $M_{N_0}$ starting from an unconstrained thermal equilibrium,  $P^\mathrm{PQ} (M_{N_0})$ reads:
\begin{equation}
	P^\mathrm{PQ} (M_{N_0}) \propto \binom{N_0}{\frac{N_0+M_{N_0}}{2}}  
	e^{\frac{j}{N_0}M_{N_0}^2 }
\end{equation}
We can extend this property to systems in which a part of it has already been quenched with a specific magnetization.
\paragraph{
	Recycled Quenching (RQ) \cite{PQ-CM-KS-2022}}:
Moreover, we have studied the effect of an ``annealing'' by which we relax a randomly chosen spin among the fixed ones during PQ. When we repeat the cycle of random unquenching and random quenching, the distribution of the fixed magnetization $M_T$ of $T$ quenched spins converges asymptotically to the distribution which we would have by applying PQ up to $T$ spins starting from unconstrained equilibrium.

\section{PQ in Markovian models}\label{sec:PQ_in_markov_models}
In this section, we limit our consideration to the Markovian evolution models and investigate how the canonicity of the statistics plays a role.
\subsection{Canonicity upon PQ in two-story ensemble}\label{subsec:two-story}
In the {\it Note added in proof} of \cite{PQ-CM-KS-2022}, we predicted in the forthcoming paper a further description about the origin of canonical distribution of the final quenched magnetization, $M_{T=N_0},$ in fact with the implicit assumption of Markovian dynamics.
The main questions are:\\
(i) A problem of combinatorics (Sec. \ref{subsub:combinatorial}):  How the quenched ensemble is compatible with the canonical statistics upon the consecutive quenching operations where the unquenched spins are in constrained equilibrium ?\\
(ii) A problem of dynamics (Sec.  \ref{subsub:dynamical}): At the level of discrete spins - and even more microscopic - how is the operation of quenching compatible with the reversible evolution, given the apparent Deborah number, that is, the dimensionless ratio of the relaxation time of the system and the observation period, exceeding unity? Indeed, the quenching process physically implies rendering towards zero the transition rate for the flipping of the spin in question. 

\subsubsection{Combinatorial approach}\label{subsub:combinatorial}
First, we introduce the notion of {\it two-story ensemble}, the way of characterizing the 
statistics of $N_0$ spins, which is convenient for the PQ.
We separate those $N_0$ spins into two groups, $\{s_1,\ldots,s_T\}$, and the remainder, $\{s_{T+1},\ldots,s_{N_0}\}$ (keeping the quenched/free spins distinction in mind) and we introduce the sub-totals of spins through $M_T= \sum_{i=1}^{T} s_i$ and $\mu_T =\sum_{j=T+1}^{N_0} s_j.$ 
The joint probability $P_{QF}(M_T,\mu_T)$ satisfies  $P_{QF}(M_T,\mu_T)=P_{F|Q}(\mu_T|M_T)P_Q(M_T),$ where $P_Q(M_T)$  is the marginal and $P_{F|Q}(\mu_T|M_T)$ is the conditional probability.\footnote{For the simplicity of notations we suppressed 
the index $T$ as the number of quenched spins. For example, it is understood that $P_{QF}(M_T,\mu_T)$ is for the $T$ quenched spins.}
We interpret this identity in the way that $P_Q(M_T)$ characterizes the {\it families} of spin configurations in the quenched part, $\{s_1,\ldots,s_T\},$
while $P_{F|Q}(\mu_T|M_T)$ reflects the sub-ensemble of the spin configuration, $\{s_{T+1},\ldots,s_{N_0}\},$ in each family member.
The configurations in the same family are realized ergodically, while those belonging to a distinct quenched family are non-ergodic in the two-story ensemble.
(In our model on a complete graph, we replaced $\{s_1,\ldots,s_T\}$ by $M_T$ as a collective tag of the family.)

\begin{figure}
\includegraphics[width = \linewidth]{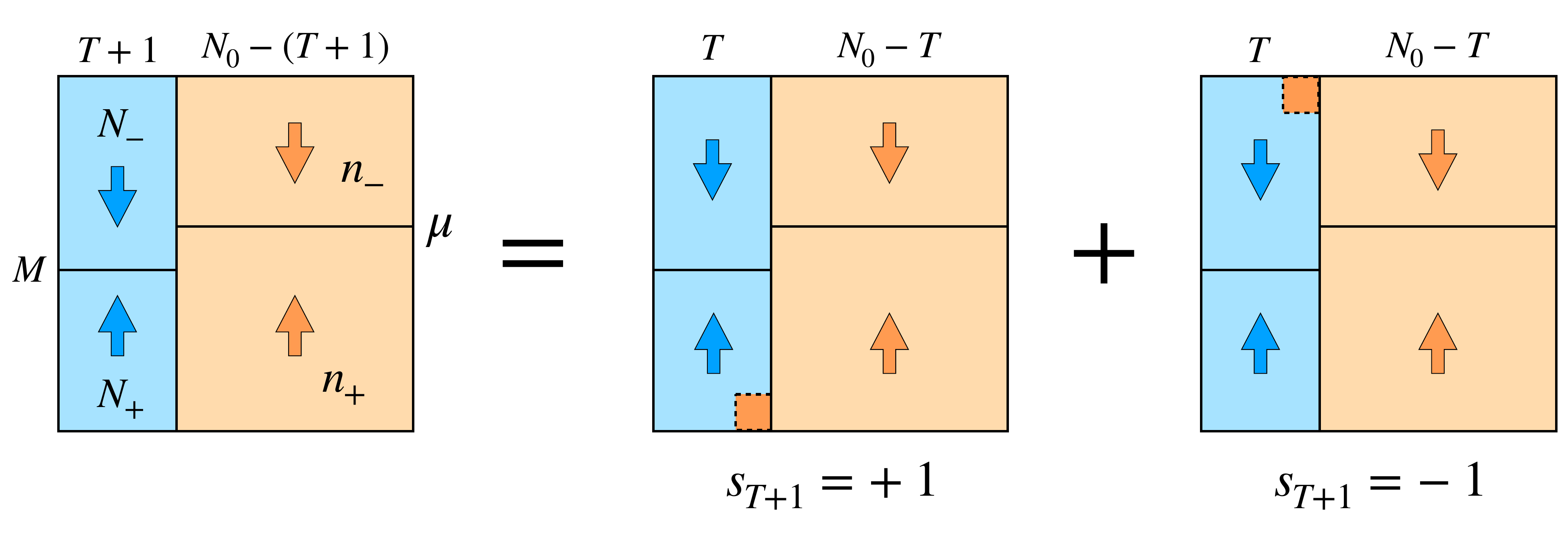}
\caption{Schematic representation of the process of updating the two-story ensemble. The blocks represent the partition of the spins into the quenched part (the left column in blue) and the free part (the right column in red). According to the sign of spins, each column is subdivided: 
	$N_+ +N_-=T+1$ and $n_+ + n_-=N_0-(T+1)$ while $N_+ -N_-=M(=M_{T+1})$ and $n_+ - n_- = \mu (=\mu_{T+1}).$ }
\label{fig:reindexation}
\end{figure}

The above is for a particular two-story ensemble. The different values of $T$ define the distinct constructions of two-story ensembles.
In the context of PQ, however, we introduce a particular form of connection between
a two-story ensemble $\{P_Q(M_T), P_{F|Q}(\mu_T|M_T)\}$
and its ``neighbor'' ensemble, $\{P_Q(M_{T+1}), P_{F|Q}(\mu_{T+1}|M_{T+1})\}.$ 
This connection is schematically explained in Fig.\ref{fig:reindexation},
where $N_\pm$ and $n_\pm$ denote the numbers of up/down ($\pm 1$) spins in the quenched and free parts, respectively.
All the spins are initially unquenched ($T=0$ and $M=0$) and thought to be in equilibrium without an external field. The probability for 
$\mu_0,$ denoted by $P_{F}^{(can)}(\mu_0)$ reads:
\begin{equation} \label{eq:simple-canonical}
	P^{(can)}_{F}( \mu_0)=
	\mathcal{N}_{0,0}  \binom{N_0}{\frac{N_0+\mu_0}{2}}  
	e^{\frac{j}{N_0}\mu_0^2 },
\end{equation}
where the normalization constant $\mathcal{N}_{0,0}$ is  
such that  	$ \sum_{\mu=-N_0}^{N_0\, (mod\, 2)} 
P^{(can)}_{F}( \mu_0)=1.$
We further assume that, upon the quenching of the $(T+1)$-th spin, 
the $(N_0-T)$ free spins have already been re-equilibrated under the given fixed magnetization, $M_T.$ 
In Appendix \ref{app:canonicity} we show by induction that the joint probabilities, $P_{QF}(M_T,\mu_T)$ for all $T,$ obey the canonical statistics 
if the initial weight $P_F(\mu_0)$ obeys the canonical statistics and 
that the PQ fixes the value of any one of the free spins under constrained canonical equilibrium. 
 % canonical-quench.tex
Regarding the statistics of quenched spins, the above result implies that at any stage, for example, the $T$-th stage, their magnetization $M_T$ is distributed as if the $T$ spins were randomly sampled from an equilibrium ensemble of $N_0$ spins.

{\it Relation to martingale :} 
In light of the canonicity underlying the two-story ensemble of quenched and free spins, the mechanism that allowed the ``martingality'' of $\meq_{T, M_T}\equiv \mathbb{E}[{\null} s_{T+1}|{M}_T]$ is easily understood: 
\null{While $\mathbb{E}[{\null} s_{T+1}|{M}_T]$ originally
meant the expectation of the spin $s_{T+1}$ upon quenching in the presence of the magnetization $M_T$ due to the $T$ already fixed spins},
the underlying canonicity allows us to map it to the equilibrium expectation of $s_{T+1}$ when the spins $\{s_1,\ldots, s_T\}$ have the magnetization $M_T.$ Together with the homogeneity among the free spins, $\{s_{T+1},\ldots, s_{N_0}\},$ we finally regard
$\meq_{T, M_T}$ as the canonical expectation $\mathbb{E}^{(can)}[s_{N_0} |M_T].$
The ``$M_T$-martingality'' for the latter follows directly from the tower rule applied to $m_T\equiv \mathbb{E}[{\null} z |s_1,\ldots, s_T],$
\null{ see Appendix \ref{app:tower}}, where $z$ stands for any random variable belonging to the above canonical ensemble. 
In this viewpoint, we can better understand the effect of Recycled Quenching (RQ) \cite{PQ-CM-KS-2022} mentioned at the end of Section \ref{subsec:IIB}. After applying the RQ  infinitely many times, the probability $P_Q(M_T)$ of having the quenched magnetization $M_T$ over $T$ quenched spins in fact obeys 
the canonical marginal distribution, $\sum_{\mu_T}P^{(can)}(\mu_T,M_T),$ where $P^{(can)}(\mu_T,M_T)$ is the joint canonical distribution of the spins when $T$ randomly chosen spins have the magnetization $M_T.$ 

The picture of the two-story ensemble and the underlying canonical statistics should apply to systems other than the Ising spins on a complete graph.  
See, as an example, Appendix \ref{app:potts} for the $q=3$ Potts model.
Note that the equivalence between the martingale and the local invariance \cite{PQ-CM-KS-2022} is, however, specific to the Ising spin model.

\subsubsection{Dynamical approach - Finite time reversible operation}\label{subsub:dynamical}

While the conservation of canonical structure was shown based on the statistical argument above, we here check whether the dynamics of quenching can be performed reversibly. The argument is done at the level where the spins span a discrete state space.
We will see below that, on the Glauber level the canonical equilibrium weight can be maintained under the time-dependent kinetic parameters.
In Subsection \ref{subsubsec:continuous-TN} we will come back to the dynamics problem at a more microscopic level where each spin takes continuous states in a double-well potential.

\begin{figure}
\centering
\includegraphics[width = \linewidth]{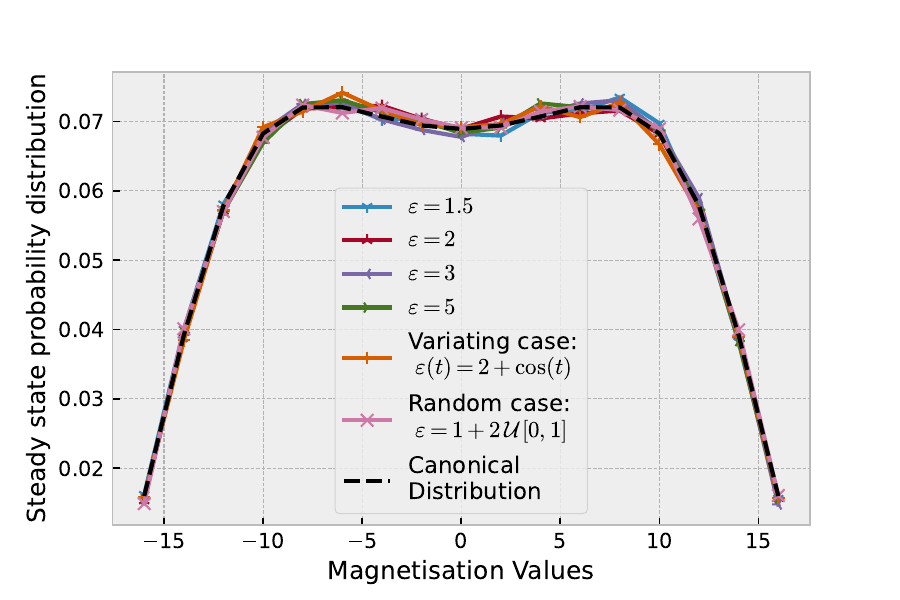}
\caption{Invariance of the steady state distribution of Glauber algorithm with different values of $\varepsilon$, either fixed with time, variating, or chosen randomly at each time step. $\mathcal{U}[0,1]$ stands for a uniform random variable over the interval $[0,1]$. } 
\label{fig:Steady_state_invariance_with_eps}
\end{figure}

Glauber's algorithm  \cite{glauber1963time, martinelli1999lectures} is a representative model of continuous-time Markovian evolution of Ising spins. In this model the flipping of the Ising spin $s_{i}$ in the presence of the interacting energy,
\begin{equation} \label{eq:defEi}
	E_{i}(t) = j \sum_{k ( \neq {i})} s_k(t), 
\end{equation}
is characterized by the transition rate of the single-spin flip:
\begin{equation}\label{eq:Glauber_flip_proba}
	P[s_{i}(t+dt) = - s_{i}(t)] = \frac{dt}{2\varepsilon_i(t)}({1 - s_{i}(t) \tanh (\beta E_{i}(t))}) ,
\end{equation}
where the characteristic time $\varepsilon_i(t)$ may depend on the time $t$.
In this context the operation of quenching spin $s_{T+1}$ amounts to render $\varepsilon_{T+1}(t)$ to $+\infty.$ 
On the other hand, we know that if the time constants $\{\varepsilon_{i}\}$ are 
static, the above algorithm can establish the canonical distribution as its steady state with whatsoever values of $\{\varepsilon_i\}.$ While the latter does not immediately imply that the quenching, or general time-dependent modulation of  $\varepsilon_{i}$'s, allows the canonicity to be kept intact against the dynamic perturbation, it is assured by the fact that the Kullback-Leibler divergence, 
\begin{equation*}
	D(P^{} \| P^{can})=-\sum_{\{s_i\} }P({\{s_i\}},t) \ln\frac{P(\{s_i\},t)}{P^{can}({\{s_i\}})}, 
\end{equation*}
is a Lyapunov functional of the Markovian evolution of $P(\{s\},t)$ whether or not $\{\varepsilon_{i}\}$ are time-dependent\footnote{In the generic inequality, $D({\sf K}P \| {\sf K}Q)\le D(P\| Q)$  for the probability vectors $P$ and $Q$ with a transfer matrix ${\sf K},$ we substitute $P=P_t,$ $Q=P^{can}$ and ${\sf K}=\bm{1}+dt\,{\sf R},$ where ${\sf R}$ is the rate matrix. Then we have $D(P_{t+dt}\|P^{can})\le D(P_t\| P^{can}).$}.
Figure \ref{fig:Steady_state_invariance_with_eps} demonstrates 
that the Glauber model keeps the canonicity regardless of the choice of characteristic times $\{\varepsilon_{i}\},$ either static or dynamic.
To summarize, on the level of discrete spin space, the quenching means to render to infinity the elementary transition time of the Glauber model, and this operation can be performed in keeping the detailed balance in the Markovian transition network.

\subsection{PQ viewed in the transition network} \label{subsec:Markov_DB}
\subsubsection{Transition network of the spin system on complete graph}
\label{subsubsec:SpinTN}
Here we aim to extend the PQ to the context of Markovian transition networks (TN).
First, we translate the PQ of a spin system in the language of TN, 
\null{ i.e., we will deal with all possible spin configurations and not just their quenched magnetization, $M_T$ ($1\le T\le N$). This \textit{fine-grained} description allows us to take into account both quenched and unquenched spins as the ``coordinates'' of TN.
	When $p_T(s_1,\ldots,s_N)$ is the fine-grained probability distribution of the spins of which $\{s_1,\ldots,s_T\}$ are quenched and the remainder,
	$\{s_{T+1},\ldots,s_N\}$ are in the constrained equilibrium state, the probability distribution of the coarse-grained random variable $\hat{M}_T\equiv \sum_{i=1}^T s_i,$ which we denote by $P(M_T),$ is given by
	\beqa \label{eq:ET-from-sis}  
	P(M_T)&=&\mathbb{E}[{\null} \delta_{\hat{M}_T,M_T}]
	\cr &:=&\!\!\!\sum_{s_1=\pm 1}\!\!\!\cdots\!\!\!\sum_{s_N=\pm 1}
	\!\!\!\delta_{\hat{M}_T,M_T} p_T(s_1,\ldots,s_N),
	\eeqa
	where $\delta_{a,b}$ ($a,b\in \Z$) 
	is the Kronecker's delta. We notice that the joint distribution $p_T(s_1,\ldots,s_N)$ can be factorized into the conditional thermal distribution, $p_T(s_{T+1},\ldots,s_N|s_1,\ldots,s_T)$ and the quenched spin distribution given as the marginal:
	$\sum_{s'_{T+1}=\pm 1}\!\cdots\!\sum_{s'_N=\pm 1} p_T(s_1,\ldots,s_T,s'_{T+1},\ldots,s'_N).$
}

Slightly more generally, we consider a system with $N$ degrees of freedom denoted by $ \{x_1, x_2, \dots, x_N \} $. The set of possible values of $x_i$ is denoted by $A_i$. For example, $A_i=  \{-1,1 \}$ for an Ising spin $s_i$ and we define $M_T:=\sum_{i=1}^T A_i$ when the first $T$ degrees of freedom are quenched. The state space $A$ then reads $A \equiv \bigotimes_{i=1}^N A_i.$
Any state $\alpha \in A$ can then be described by a set of degrees of freedom, $ \{x_1, x_2, \dots, x_N \}.$ Inversely, any variable $x_i$ is the function of the state, $x_i(\alpha).$
The transition network in $A$ is such that (i) if we exclude the simultaneous change of more than one variable, the topology of transition edges before quenching is hyper-rectangle, and (ii) if any one variable, e.g. $x_i,$ is quenched, the network is divided into two groups, losing the ergodicity. Fig \ref{fig:PQ-in-TN}(b) illustrates (i) and (ii) for three spins ($N=3$), where an initial TN graph is divided into non-connected subgraphs.

Let $\bm{\mathcal{R}}$ be the rate matrix of the master equation for the network on $A$:
\begin{equation*}
	\frac{d \vec{P}}{dt} = {\bm{\mathcal{R}}}\vec{P} , 
\end{equation*}
and let $\vec{P}^\mathrm{st}$ be the steady state distribution; 
${\bm{\mathcal{R}}}\vec{P}^\mathrm{st} = \vec{0}.$
We also introduce the net probability current from $\alpha$ to $\alpha'$ through
$$
J_{\alpha'\leftarrow\alpha} \equiv \mathcal{R}_{\alpha'\leftarrow\alpha}P_\alpha
-\mathcal{R}_{\alpha\leftarrow\alpha'}P_{\alpha'}.
$$
When the detailed balance (DB) is established for the steady state, $\vec{P}^{\,st},$
we have $J_{\alpha'\leftarrow\alpha}=0$ for all the pair of states, $(\alpha,\alpha').$

Having the progressing quenching in mind, we introduce the class-Kronecker delta, $\delta^{(T)}_{\alpha,\alpha'} (=\delta^{(T)}_{\alpha',\alpha})$ through\footnote{$\wedge_{i=1}^T\{C_i\}$ means $C_1\wedge \cdots\wedge C_T$, where $\wedge$ is the conjunction operator.}
\beq
\delta^{(T)}_{\alpha,\alpha'}=
\left\{\begin{array}{ll}
	1    &: \quad\wedge_{i=1}^T \{x_i(\alpha)=x_i(\alpha') \} \\
	0    &: \quad\mbox{otherwise}
\end{array}\right.  ,
\eeq
that is, it picks up those pair of states that belong to a subset of the same tag, $\{x_1,\ldots,x_T\}.$ 
When the progressive quenching has fixed $\{x_1,\ldots,x_T\}$ but leaves the other variables free to fluctuate, the modified rate matrix, which we denote by $\tilde{\mathcal{R}}_{T, \alpha' \leftarrow \alpha }$ is given as 
\beq
\label{eq:reatribution_MC}
\tilde{\mathcal{R}}_{T, \alpha' \leftarrow \alpha } 
= \delta^{(T)}_{\alpha,\alpha'}  \mathcal{R}_{\alpha' \leftarrow \alpha } 
\eeq
for $\alpha\neq \alpha',$ and 
$\tilde{\mathcal{R}}_{T, \alpha \leftarrow \alpha} =
- \sum_{\beta(\neq \alpha)}^{}
\tilde{\mathcal{R}}_{T,\beta \leftarrow \alpha } $ 
for the diagonal element to satisfy the normalization conditions, $\sum_{\alpha'} \tilde{\mathcal{R}}_{T,\alpha' \leftarrow \alpha} = 0$ for $\forall \alpha.$ 
Eq. (\ref{eq:reatribution_MC}) simply means the state transition is possible only when $\delta^{(T)}_{\alpha,\alpha'}=1.$ 

A simple but important observation is that if the steady state of the unquenched system, 
$\vec{P}^\mathrm{{st}},$ satisfies the detailed balance, 
then we have a trivial rewriting for every pair $(\alpha,\alpha'),$
\beqa \label{eq:zeroJremovable}
0 &=& J_{\alpha'\leftarrow\alpha} 
\cr &=&
\mathcal{R}_{\alpha'\leftarrow\alpha}P_\alpha^{\rm st} -\mathcal{R}_{\alpha\leftarrow\alpha'}P_{\alpha'}^{\rm st}
\cr &=&
\delta^{(T)}_{\alpha,\alpha'} \inRbracket{\mathcal{R}_{\alpha'\leftarrow\alpha}P_\alpha^{\rm st} - \mathcal{R}_{\alpha\leftarrow\alpha'}P_{\alpha'}^{\rm st}}
\cr &=&
\tilde{\mathcal{R}}_{\alpha'\leftarrow\alpha}P_\alpha^{\rm st} -\tilde{\mathcal{R}}_{\alpha\leftarrow\alpha'}P_{\alpha'}^{\rm st}.
\eeqa
This means that $\vec{P}_{}^{\,\rm st}$ satisfies also the DB condition for the 
quenched system.
The steady states of $\widetilde{{\bm{\mathcal{R}}}}$ are in general not unique because of the {\it broken ergodicity}  (see Fig \ref{fig:PQ-in-TN}(b)). Nevertheless, the canonical distribution, $\vec{P}^\mathrm{{st}},$ is among the possible steady states.

\subsubsection{Comments on the continuous state description}
\label{subsubsec:continuous-TN}

The fact that the canonical equilibrium weight can be maintained under the time-dependent kinetic parameters at the level of Glauber dynamics (Sec.\ref{subsub:dynamical}) does not automatically assure the existence of a microscopic continuous-state model that reproduces the persistence of the canonicity. We show below that, on the microscopic level, we can quench a single spin under a fixed external field in keeping the detailed balance, i.e. the canonicity, but that its extension to multi-spin systems seems to be at best approximative, which we argue using the landscape argument.
	
\begin{figure}
	\centering
	\includegraphics[width = \linewidth]{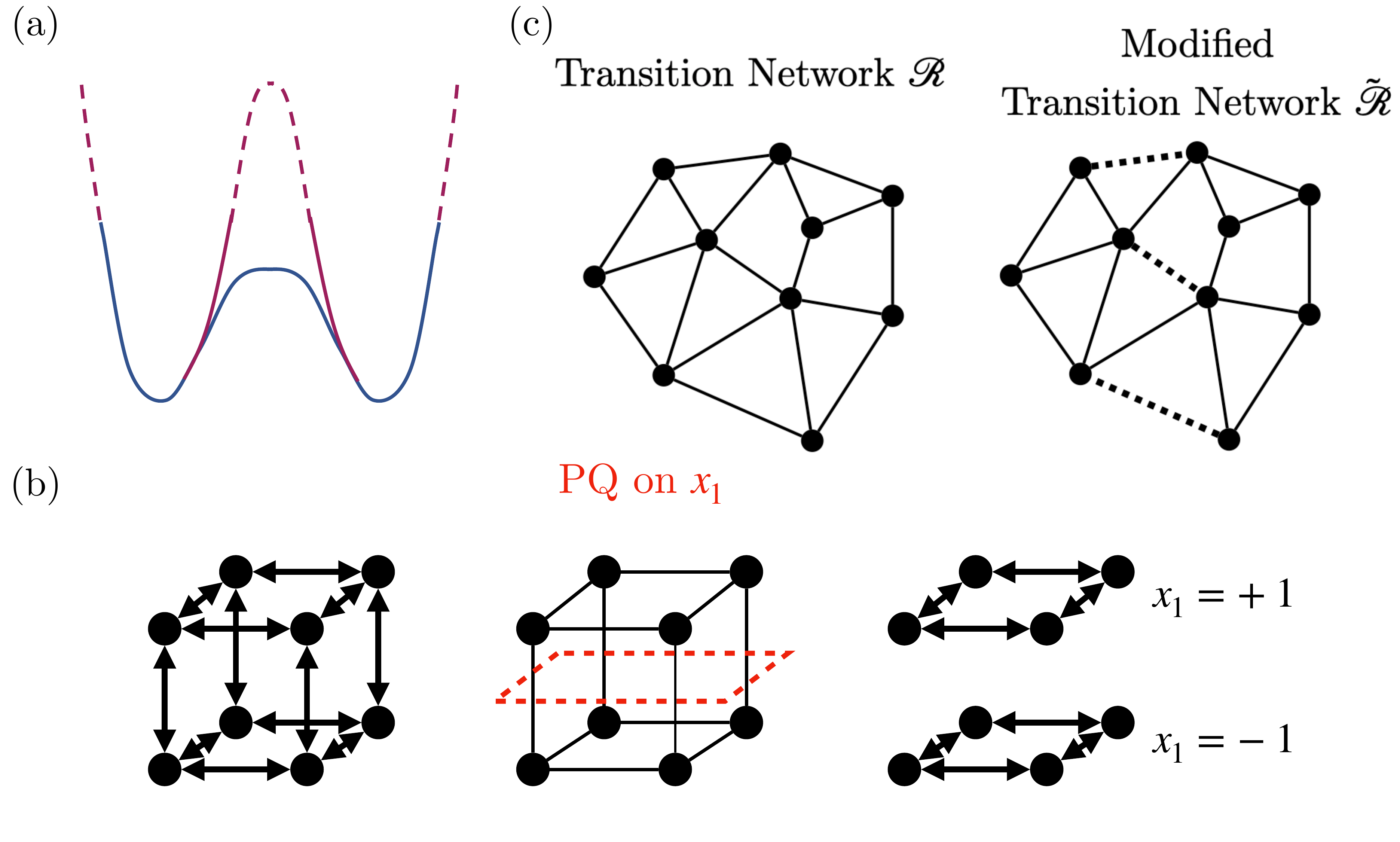}
	\caption{Different viewpoints of Progressive Quenching.
		{(a)} A double-well potential as a microscopic model of single-spin quenching. The dashed parts of the curves are those inaccessible by thermal activation with the experimental time.
		{(b)} Schematic illustration of a cubic Markov transition network modulated by PQ (in the case of three Ising spins for example). After quenching the first degree of freedom denoted by $x_1$, the cubic network is separated into two square disconnected subnetworks.
		{(c)} Schematic illustration of a network transformation. Some edges whose
		net probability flow has been zero are removed (dashed line). }
	\label{fig:PQ-in-TN}
\end{figure}

In well-known modeling of the bit-memory analyzed by Landauer \cite{landauer}, the individual classical spin (or a binary bit) under a static bias field was visualized as a state point in a double-well potential, see Fig.\ref{fig:PQ-in-TN}(a). The quenching of a spin corresponds to the raising to infinity of the barrier separating the double-well.
Though being oversimplified it will be instructive to quantify the irreversibility associated with the manipulation of the barrier height of the double-well potential.
The partial entropy production introduced by Shiraishi \cite{shiraishi2016measurement, shiraishi2016_arxiv_Partial_entropy_prod, shiraishiPRL2016_Trade_off_Power_eff} may be fitted for this purpose. 
If we approximately discretize the coordinate $x$ of the double-well potential (Figure \ref{fig:PQ-in-TN}(a)), the partial entropy production associated with the (nearby) transitions $x'\to x$ denoted by  $\dot{S}_{x,x'}$ reads
\beq \label{eq:ShiraishiSdot}
\dot{S}_{x,x'}= R_{x x'}p_{x'}
\ln\frac{R_{x x'}p_{x'}}{{R}_{{x}' {x}}p_{x}}
+ {R}_{{x}' {x}}p_{x}-R_{x x'}p_{x'},
\eeq
where $p_x$ is the probability and $R_{x',x}$ is the transition rate from $x$ to $x'$ and we assumed that the time-reversed state of $x$ is $x$ itself.
When the potential is modified sufficiently slowly relative to the microscopic time scale, the probability flows, $R_{x x'}p_{x'}-{R}_{{x}' {x}}p_{x},$ with $x$ and $x'$ within the same valley, remain effectively zero through the detailed balance, with the only exception around the barrier top.
In Appendix \ref{app:Shiraishi}, we demonstrate that, by focusing on the barrier top, this framework gives the famous Landauer's entropic loss by $\ln 2$ upon the erasure process of a bit memory.
By contrast, in the present context, the quenching of a spin
is made so that the DB is observed {\it including} at the vicinity of the barrier top. Then the local entropy production $\dot{S}_{x,x'}$ in (\ref{eq:ShiraishiSdot})  vanishes everywhere.

With the interacting $N(>1)$-spin system, however, the above argument should be lifted to a high-dimensional phase space, $\R^N.$ 
We suppose a free energy landscape made by the potential,
$\sum_i \phi(x_i,\lambda_i) - j\sum_{i<j} x_i x_j,$ where $x_i$ is the continuous coordinate of the $i$-th spin and the individual potential $\phi(x_i,\lambda_i)$ has minima at $x_i=\pm 1.$ The initial detailed-balanced statistical state implies the vanishing probability flow 
on this free energy landscape \cite{Lebowitz55}.
We have seen in \S\ref{subsubsec:TN-Mtgl} that quenching of a spin in the presence of the other (unquenched) spins implies to remove $2^{N-1}$ edges of the transition network (see Fig.\ref{fig:PQ-in-TN}(b)). 
As for the continuous-state version, if the $i$-th coordinate $x_i$ is quenched through a single function $\phi(x_i,\lambda_i)$ by operating the parameter $\lambda_i,$ it causes the {\it simultaneous} raising of $2^{N-1}$ barriers. 
Since the local landscape in the neighborhood of each of the $2^{N-1}$ barriers reflects a particular configuration of the remaining $(N-1)$ spins, the raising of those barriers through a single function $\phi(x_i,\lambda_i)$ should inevitably bring about a non-zero probability flow across several barriers though it can be quantitatively small. 
It is the reason why we think the extension of reversible quenching for a multi-spin system seems to be at best approximative at the microscopic level.

\subsection{PQ of Markovian transition network without detailed balance}\label{subsec:MarkovNonDB}
\subsubsection{Removal of an edge without modifying steady-state probabilities}
When the states space $A$ is not a product space corresponding to multiple degrees of freedom of the system, we may still consider the action of quenching as the elimination of a part of bidirectional edges from the transition network (TN). 
If the detailed balance (DB) condition is not globally satisfied, the removal of bidirectional edges in a TN generally causes the modification of its steady-state distribution.
\begin{figure}
\centering
\includegraphics[width = 0.8\linewidth]{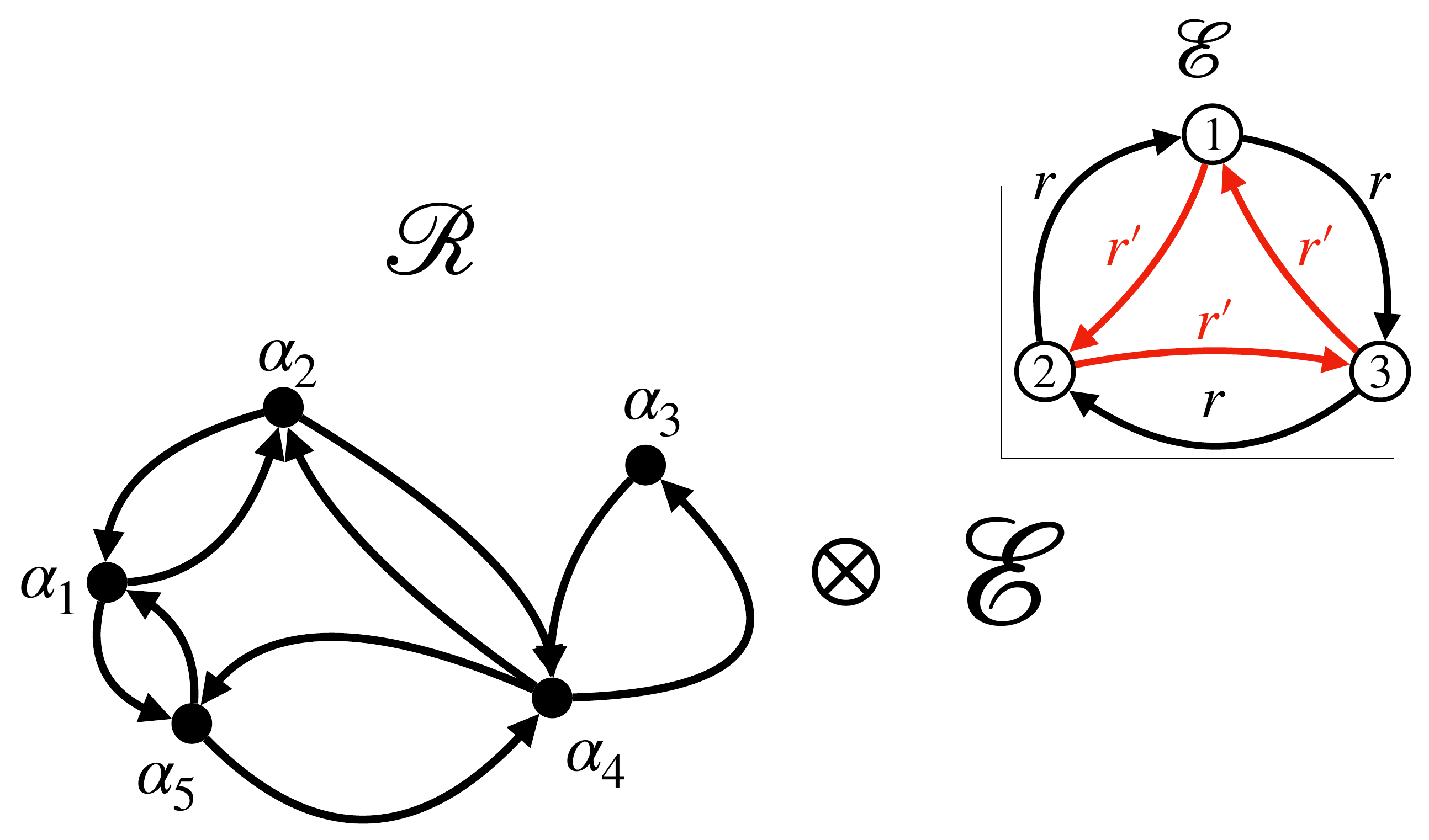}
\caption{
An example of a two-layered transition network whose steady state does not verify the detailed balance for any pair of states but yet has the possibility of a ``quench'' leaving the stationary probabilities intact. See the main text (\ref{subsec:MarkovNonDB}.1) for details.
	(inset) A simple example of a stationary Markov chain $\mathcal{E}$
without detailed balance. If $r \neq r'$, there is a non-zero probability flux, and cutting a pair of antiparallel arrows will cause a change in the stationary distribution. 
}
\label{{fig:two-layered}}
\end{figure}
The inset of Fig \ref{{fig:two-layered}} only shows a simple example where the stationary state has a circulation of probability. Before ``quenching'' the stationary probability on the three states is $\{p_1,p_2,p_3\}= \left\{\frac{1}{3},\frac{1}{3},\frac{1}{3} \right\}.$ When we remove the edges between the states $1$ and $2,$ the stationary probability becomes
${\left(r^2+r r'+ r^2 \right)}^{-1} \left\{r'^2, r^2, r r' \right\}.$ 
It is only when $r=r'$ that the detailed balance holds globally, and the stationary distribution remains unchanged by this operation.

When we consider the general TN and ask when the removal of bidirectional edges leaves the steady state probability intact, a rule of thumb is as follows:\\
When a pair of states, $(\alpha, \alpha')$ have realized the vanishing net probability flow, $J_{\alpha'\leftarrow \alpha}=0,$ we can eliminate simultaneously $\mathcal{R}_{\alpha'\leftarrow\alpha}$ and $\mathcal{R}_{\alpha\leftarrow\alpha'}$ without perturbing the stationary distribution.
The demonstration follows the idea of (\ref{eq:zeroJremovable}) above. We suppose that the initial TN has a steady state $\vec{P}^{\,\rm (st)}.$  We denote by $\chi_Q$ all those pairs of states for which the net probability flow vanishes, i.e., 
\begin{equation}
\chi_Q \equiv \left\{(\alpha, \alpha') \,| \,
\mathcal{R}_{\alpha^{\prime}\leftarrow\alpha} P_\alpha^{\,\rm (st)} - \mathcal{R}_{\alpha\leftarrow\alpha'}P_{\alpha'}^{\,\rm (st)}=0
\right\}.
\end{equation}
Here $\chi_Q$'s suffix $Q$ stands for ``quenchable''. We introduce the ``optional''-Kronecker delta, $\delta^{(Q)}_{\alpha,\alpha'}$ though
\beq
\delta^{(Q)}_{\alpha,\alpha'}=\delta^{(Q)}_{\alpha',\alpha}=
\left\{\begin{array}{ll}
\mbox{1 or 0\;(optional)}   &: \quad(\alpha,\alpha')\in \chi_Q \\
1    &: \quad\mbox{otherwise}
\end{array}\right.  ,
\eeq
that is, $\delta^{(Q)}_{}$ can vanish only for the pairs whose net steady probability flow is zero. We then ``quench'' the original TN according to the ``optional''-Kronecker delta: 
\begin{equation}\label{eq:Rmodif}
\tilde{\mathcal{R}}_{\alpha^{\prime} \leftarrow \alpha}\equiv
\delta^{(Q)}_{\alpha,\alpha'}
\mathcal{R}_{\alpha^{\prime}\leftarrow\alpha}.
\end{equation}
We can check that the ``quenched'' TN still has $\vec{P}^{\,\rm (st)}$ as its stationary state. In fact, for every $\alpha'$ 
\begin{align}
&\sum_{\alpha}( \tilde{\mathcal{R}}_{\alpha^{\prime}\leftarrow\alpha} P_\alpha^{\,\rm (st)} - \tilde{\mathcal{R}}_{\alpha\leftarrow\alpha'}P_{\alpha'}^{\,\rm (st)}) \notag \\
&=\sum_{\alpha} \delta^{(Q)}_{\alpha,\alpha'}
( {\mathcal{R}}_{\alpha^{\prime}\leftarrow\alpha} P_\alpha^{\,\rm (st)} - {\mathcal{R}}_{\alpha\leftarrow\alpha'}P_{\alpha'}^{\,\rm (st)})\notag \\
&=\sum_{\alpha} 
( {\mathcal{R}}_{\alpha^{\prime}\leftarrow\alpha} P_\alpha^{\,\rm (st)} - {\mathcal{R}}_{\alpha\leftarrow\alpha'}P_{\alpha'}^{\,\rm (st)})\notag \\
&= 0.
\end{align}
Here, to go to the third line, we have used the fact that whenever the pair $(\alpha, \alpha')$ is $\not\in \chi_Q,$ we have $\delta^{(Q)}_{\alpha,\alpha'}=1$ by definition. The last equality is the stationary condition for the original TN.
The central part of Fig.~\ref{{fig:two-layered}} gives an example in which the TN does not have a global detailed balance, but the ``quenching'' of TN is possible.
The system has two layers, $\mathcal{R}$ and $\mathcal{E}.$ 
The former layer has the Glauber dynamics allowing detailed balance
among $\{\alpha_1,\alpha_2,\alpha_3,\alpha_4,\alpha_5\}.$
The latter layer $\mathcal{E}$ undergoes the stochastic circulation among
\null{ $\{1,2,3\},$} see the inset.
We assume that the \null{ three} values $\varepsilon_k$ \null{ ($k=1,\ldots,3$)} are the values of the global time-constant of the Glauber dynamics for the first layer $\mathcal{R}.$
Then, we can quench the bidirectional edges for any pairs of nodes on this layer.
\paragraph{Remark}: 
The modification of the transition rates, 
${\mathcal{R}}_{ \alpha' \leftarrow \alpha } \mapsto \tilde{\mathcal{R}}_{ \alpha' \leftarrow \alpha } $ and 
${\mathcal{R}}_{ \alpha \leftarrow \alpha' } \mapsto \tilde{\mathcal{R}}_{ \alpha \leftarrow \alpha' } $
should be realized {\it pairwise} simultaneously, either instantaneously or gradually, but in keeping the ratio 
$ \tilde{\mathcal{R}}_{\alpha^{\prime}\leftarrow\alpha}(t)/ \tilde{\mathcal{R}}_{\alpha\leftarrow\alpha'}(t)$ constant so as to maintain
the flow-free condition, $( \tilde{\mathcal{R}}_{\alpha^{\prime}\leftarrow\alpha}(t) P_\alpha^{\,\rm (st)} - \tilde{\mathcal{R}}_{\alpha\leftarrow\alpha'}(t)P_{\alpha'}^{\,\rm (st)})=0.$

\subsubsection{Extension of PQ and its martingale to Transition Network}\label{subsubsec:TN-Mtgl}

In fact, the martingale is more basic than the canonicity when we extend the PQ to the generic Markovian networks. The logic is as follows: 
The operation of PQ amounts to stretching to infinity the microscopic response time of individual spin ($\varepsilon_i$ for the $i$-th spin for the Glauber model, \S\ref{subsub:dynamical}). In the extension to general Markovian transition networks (TN), we can progressively remove the bidirectional transition edges in a prefixed manner until all the individual nodes become isolated.

Now the general condition of the martingale of the tower-rule type is that there is the conditional expectation of a {\it common random variable,} say $z,$ which is associated with a growing filtration, i.e.., a sequence of cumulative conditions in the form, 
$\mathbb{E}[{\null} z],\mathbb{E}[{\null} z|X_1],\mathbb{E}[{\null} z|X_1, X_2], \ldots$ In the original PQ of spins, while the sequence of the fixed spin naturally gave a growing filtration (\S\ref{subsubsec:SpinTN}), we resorted to the canonicity of $M_T$ so that the $\meq_T$ meets the general condition of the martingale of the tower-rule type just mentioned. When we extend PQ for discrete-state Markovian TN, we may find a new sequence of conditional expectations without resorting to canonicity or detailed balance.

In Appendix \ref{app:TNmartingale}, which is also dedicated to this question,
we show a concrete construction of a growing filtration specified by the history, $\{X_1,\ldots, X_t\}\in (-1,0,1)^t,$ through the removal of the bidirectional edges one after another, where enough (physical) time is allowed between the consecutive quenches so that the probability flow becomes stationary in between.
There, the quenching of a spin in the original PQ is replaced by the {\it sub-division of ergodic islands} in the TN but no detailed balance on the TN is required. By choosing as $z$ the indicator function for reaching a particular node at the end, just as a trivial example, we constructed a martingale process.

\section{PQ in Non-Markovian models}\label{sec:PQ_Non_Markov}
The previous results are valid only for Markovian systems. Our study of Progressive Quenching is now extended to non-Markovian systems, whether the detailed balance (DB) is verified (Section \ref{subsec:nM-DB}) or not (Section \ref{sec:Choi_ Huberman}). The examples given are the Ising spin systems studied above but with memory effects.

\subsection{System with hidden spins satisfying detailed balance} \label{subsec:nM-DB}
\subsubsection{Model, effective coupling and DB}\label{sub2sec:recall}
In this part, we recall two known aspects of non-Markovian processes
through the case studies under a simple setup.
As a model we consider a chain of $N$ ``visible'' spins $\{s_i\}$ with ferromagnetic nearest-neighbor coupling $J.$ We also suppose that the neighboring spin pairs, say $s_i$ and $s_{i+1}$ share a ``hidden" spin
$\sigma_{i+\inv{2}}$, through the coupling
$K$. 
Fig.\ref{fig:non-Markov-deco}(a) shows the case of a closed chain with three visible spins and three hidden ones.
The energy of the entire system reads
\beq \label{eq:moselSsigma}
\mathcal{E} = -\sum_{i=1}^N Js_i s_{i+1} - K \sum_{i=1}^N (s_i+s_{i+1}) \sigma_{i+\inv{2}},
\eeq
where $s_{N+1}\equiv s_1.$ 
After taking the sub-trace over $\sigma$'s, the effective energy 
$\tilde{E}_s$ and the effective partition function $\tilde{\mathcal{Z}} $ read: 
\begin{equation}\label{eq:Equivalent_Partition_function}
\tilde{E}_s = -\sum_{i=1}^N\tilde{J}s_i s_{i+1}, \quad
\tilde{\mathcal{Z}} =\sum_{\{s_i\}} e^{-\beta \tilde{E}_s},
\end{equation}
where the effective, temperature-dependent, coupling constant $\tilde{J}$ is
\begin{equation}\label{eq:J_tilde}
\tilde{J}\equiv J+{(2\beta)^{-1}}\ln \cosh(2\beta K).
\end{equation}
The visible spins, therefore, follow the canonical statistics with the apparent coupling $\tilde{J}$ as long as the single-time statistics are concerned.\footnote{Note that, even without the quenching of spins, the hidden spins generally modify the statistical weight of the visible part from the typical form of (combinatorial factor)$\times$(a Boltzmann factor) like (\ref{eq:simple-canonical}), while the whole ensemble obeys the canonical statistics. The present example is an exception. To be general we use the term canonical \textit{statistics} instead of the canonical distribution}
If the whole system evolves by a Markovian dynamics such as the Glauber model, the observer who has only access to the visible spins $\{s_i\}$ finds its non-Markovian evolution. The non-Markovian nature in a simple case is demonstrated in Appendix \ref{subsec:Non_Markov_of_hidden_spins}.
About the visible spins, there is no more instantaneous detailed balance (DB). Nevertheless, if the whole system $\{s,\sigma\}$ obeys  a Markovian evolution with DB,  the visible spins still satisfy a \textit{trajectory-wise} detailed balance: 
\begin{equation} \label{eq:DB_for_obs_spins}
\mathbb{P}([\{s_i(t)\}_{i=1}^N]_{t=0}^T)=\mathbb{P}([\{s_i(t)\}_{i=0}^N]{^*}_{t=0}^T),
\end{equation}
where $[\{s_i(t)\}_{i=0}^N]{^*}_{t=0}^T$ denotes the time reversal of the forward trajectory, 
$[\{s_i(t)\}_{i=1}^N]_{t=0}^T.$
The derivation of Eq.(\ref{eq:DB_for_obs_spins}) is given in Appendix \ref{an:Trajectory_DB}.

\subsubsection{Effects of the PQ}
\begin{figure*}
	\centering
	\includegraphics[width = 0.99 \linewidth]{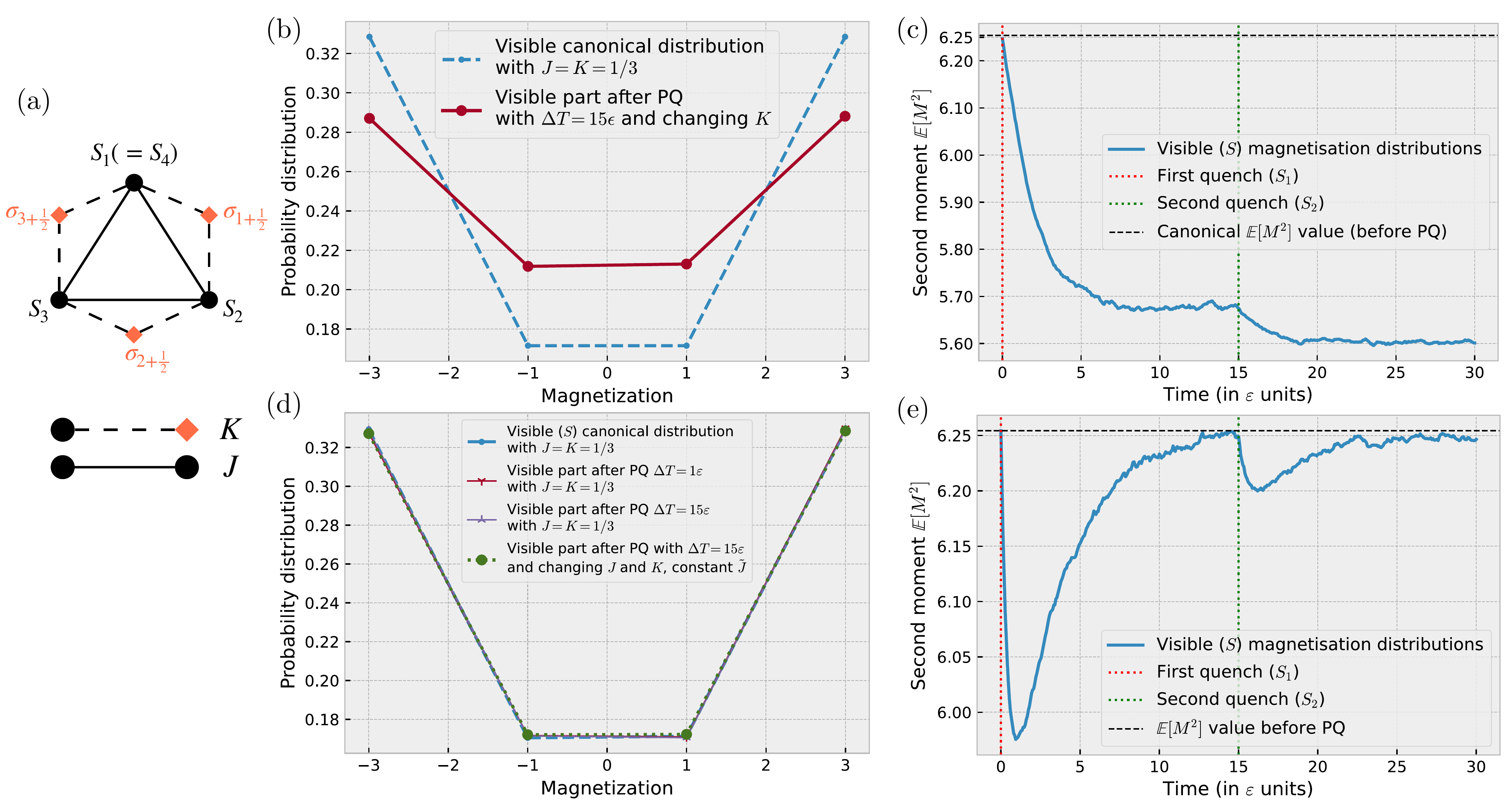}
	\caption{
		{(a)} Model of non-Markovian spin system consisting of the visible ($S_i$) and hidden ($\sigma_{i+\inv{2}}$) spins, see Eq.(\ref{eq:moselSsigma}).
		In any case studied below, the initial state obeys the canonical statistics.
		\null{\, 
			{(b)} Probability distribution of the visible spins $M=S_1+S_2+S_3$ after
			all the visible spins have been fixed. The solid red curve corresponds to the protocol in which $K$ is decreased by half at each quench, starting from $J=K=1/3.$
			For comparison, the blue dashed curve denotes the canonical equilibrium.
			(c) Transient process corresponding to relaxation dynamics when $K$ is decreased by $50 \%$ at each quench, as pictured in (b) by the red solid curve. The second moment $\mathbb{E}[{\null}  (M)^2]$ (cf. $\mathbb{E}[{\null} M]=0$) of the whole visible magnetization, $M=S_1+S_2+S_3,$ is plotted against the scaled time, $t/\varepsilon,$ 
			where the first [second] quenching are done at $t/\varepsilon=0$ [at $\Delta T/\varepsilon=15$], respectively, after each quenching.
			(d) In the same plane as (b), for the solid curves $J$ and $K$ are kept at $(1/3)(\kT)$ with the interval between consecutive quench being $\Delta T/\varepsilon = 0$(blue), $1$(red) and $15$(violet), respectively. The \null{dashed (green) curve} corresponds to variating values of $J$ and $K$ so that the effective $\tilde{J}$ remains constant - see the main text for the detailed protocol.
			{(e)} In the same plane as (c), the transient processes corresponding to the protocol for the \underline{dashed curve} in (d) are shown. 
			\null{All the data are obtained over $5.10^5$ sample trajectories. }
	}}
	\label{fig:non-Markov-deco}
\end{figure*}
\paragraph{Generic case:}
In principle, the quenching of a visible spin can accompany any actions on the hidden part. Therefore, even though the system starts with the canonical statistics, the PQ generally alters the ensemble. 
Upon quenching we may either impose the value of any hidden spin $\sigma_{i+\inv{2}}$ or change the value of the hidden coupling $K.$ 
Those hidden changes will not only drive the system transiently out of equilibrium but also alter the stationary ensemble of the visible spins being different from the original canonical statistics.
Figure \ref{fig:non-Markov-deco}(\null {b}) illustrates such a case, when the hidden coupling constant $K$ is decreased by $50\%$ after each quench, while $J$ is kept constant.
Figure \ref{fig:non-Markov-deco}(\null {c}) shows the relaxation of the second moment $\mathbb{E}[M^2]$ of the visible magnetization distribution towards the steady-state with the previous protocol.\footnote{The normalized probabilities for the three spins in the absence of an external field can always be interpreted as a canonical one of an effective temperature. The quenching makes the latter temperature different from that of the real one.}

\paragraph{Case of unbroken canonicity upon PQ:}
We take up again the non-Markovian model shown in Fig.~\ref{fig:non-Markov-deco}(a), with the energy given by Eq.(\ref{eq:moselSsigma}) wherein the steady state the detailed balance (DB) holds. As the first case of unbroken canonicity, 
a Glauber algorithm is used to simulate the dynamics of the whole system, but we progressively quench exclusively the visible spins while the hidden variables remain intact.
According to Sec \ref{subsec:Markov_DB}, the PQ of that system, in particular the selective quenching of visible spins, should not modify the distribution as the two-story ensemble. Fig.~\ref{fig:non-Markov-deco}(\null {d}) (thick curves) verifies this idea, where the probability distribution of the (visible) magnetization, $M\equiv S_1+S_2+S_3,$ after all these spins have been quenched. Here, the quenching of visible spins is progressively done with a regular (dimensionless) interval, $\Delta T/\varepsilon =0, 1$ and $15$ (solid curves), where $\Delta T/\varepsilon=0$ is equivalent to the snapshot of the equilibrium ensemble before quench. The distributions are independent of the interval $\Delta T/\varepsilon.$ 

We also examined another \textit{ad hoc} protocol in which the canonicity of the visible spins remains unbroken. This time the fixation of visible spin accompanies the modification of the coupling parameters, $J$ and $K.$ 
At every quenching, the value of $J$ is reduced by 50\% whole that of $K$ is incremented so that the effective coupling $\tilde{J}$ of Eq.(\ref{eq:J_tilde}) remains unchanged. 
When we monitor the magnetization of visible spins, $M \equiv S_1+S_2+S_3,$ its distribution after sufficient interval $\Delta T/\varepsilon$
recovers the canonical one, by construction (Fig.\ref{fig:non-Markov-deco}(\null {d}): dotted curve).
Nevertheless, there is a visible transient before the equilibration in $M,$ which we monitor through its variance $\mathbb{E}[{\null}  M^2],$ see Fig.\ref{fig:non-Markov-deco}(\null {e})  (cf. $\mathbb{E}[{\null}  M]=0$).
The fluctuations of $M$ are transiently attenuated as a fast response to the reduction of $J,$ then it recovers the canonical level (horizontal dotted line) gradually due to the compensatory increment of $K$.

\subsection{System with delayed interactions breaking detailed balance (Choi-Huberman model)} \label{sec:Choi_ Huberman}
We have seen in the previous subsection (Sec.\ref{subsec:nM-DB}) that the conservation of the canonicity upon the PQ requires a Markovian evolution rule, in addition to the detailed balance in the starting steady state. In the last part of this paper we study the effect of PQ on the non-Markovian system whose steady state has been broken from the beginning, i.e., before quenching the system's degrees of freedom, to have a better understanding of the PQ.
\subsubsection{Original Choi-Huberman model and its steady state}
The starting model is the one introduced by Choi and Huberman in 1985 \cite{choi1985collective}.
\begin{figure*}[t]
	\centering
	\includegraphics[width = 0.95 \linewidth]{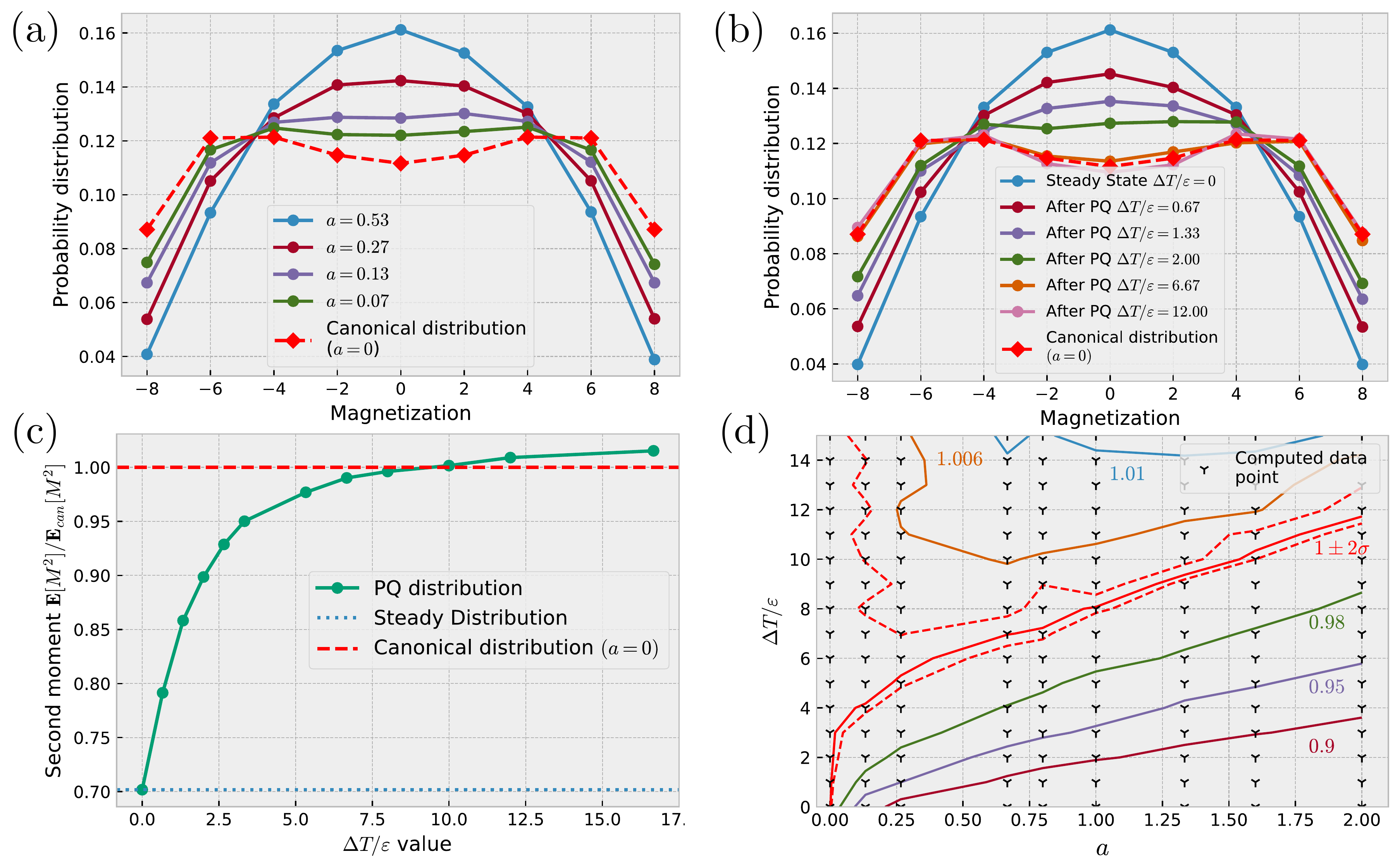}
	\caption{
		Analyses of the distribution of the total magnetization, $M=\sum_{i=1}^N s_i,$ in the Choi-Huberman model and its PQ with eight spins ($N=8$).
		(a) Plot of the steady state distribution for different values of dimensionless delay $a ={\tau}/{\varepsilon}.$ 
		The canonical distribution corresponds to the $a=0$ case.
		{(b)} Distribution of the magnetization after the Progressive Quenching has been completed with different values of $\Delta T/\varepsilon.$ The steady-state distribution, as well as the canonical distributions, are also plotted for comparison. 
		{(c)} Plot of the standardized second moment of the magnetization after the PQ, $\mathbb{E}[{\null} M^2]/\mathbb{E}^{(0)}[M^2],$ versus the time interval parameter, $\Delta T/\varepsilon,$ where 
		$a=1.07$ is kept the same as (b).
		The levels of the steady state ($\Delta T/\varepsilon=0$) and of the canonical case ($a=0$) are also shown by dashed horizontal lines. The $2\sigma$ error bars are smaller than the dots indicating calculated points.
		{(d)} Contour plot of the standardized mean square of the final quenched magnetization, $\mathbb{E}[{\null} M^2]/\mathbb{E}^{(0)}[M^2],$ on the $(a,\Delta T/\varepsilon)$ plane. Each point corresponds to a $\mathbb{E}[M^2]$ value computed over $5\times 10^5$ samples, and the contours are calculated using the ContourPy library.
		The pair of dashed contours above and below the contour of value 1 (solid red curve) corresponds to the  $1\pm 2\sigma$ values. They delimit the confidence zone where the actual contour of value 1 might be.
	}
	\label{fig:non-Markov-CH}
\end{figure*}
In their model, the interactions between spins have a delay $\tau$, i.e. each spin at time $t$ "sees" the other spins at time $(t-\tau).$ The probability of flipping of the spin $s_{i}$ reads:
\begin{equation} \label{eq:CH-dt}
	P[s_{i}(t+dt) = - s_{i}(t)]= \frac{dt}{2\varepsilon}\left[{1 - s_{i}(t) \tanh (\beta E_{i}(t-\tau))}\right] 
\end{equation}
with $E_{i}$ being defined by Eq.(\ref{eq:defEi}).
Except for the limit of the Glauber model \cite{glauber1963time} with $\tau=0,$ the steady state distribution should break the detailed balance because $\tau>0$ invalidates the time-reversal symmetry. We characterize the irreversibility by the non-dimensionalized parameter, $a \equiv \null{\tau}/{\varepsilon}.$ 
Throughout this section (\ref{sec:Choi_ Huberman}), the numerical simulation based on (\ref{eq:CH-dt}) is done with the time mesh $dt/\varepsilon=0.1$ for $a\ge 1$ and 
$dt=0.05$ for $a<1.$
In Appendix \ref{an:22CH}, we show analytically that the steady state depends on the kinetic parameter, $\varepsilon$ (via $a$), for the system with two spins.
We recall that the canonical equilibrium, i.e. $a=0,$ is independent of $\varepsilon.$
Numerically, we show in Fig.\ref{fig:non-Markov-CH}(a) how the steady state distribution 
of the system with eight spins depends on the irreversibility parameter $a.$
We see that the larger the value of $a,$ the more paramagnetic (unimodal) the system behaves as compared with the bimodal distribution with Markovian limit, $a=0.$
Intuitively, when the delay $\tau$ is augmented, the cooperative fluctuations among the spins are lessened. 
% although it is not completely hindered, which depends on $(\Delta T)/\tau.$

\subsubsection{Effects of the PQ of the Choi-Huberman model}
If we introduce the PQ in the above model of Choi-Huberman, what effect should we expect? 
First, we studied how the two-story distribution of the total magnetization evolves as a function of the number of quenched spins. The irreversibility parameter $a$ is kept at $1.07$ where 
the intact distribution is unimodal (see Fig.\ref{fig:non-Markov-CH}(a)).
We have given a large enough time interval $\Delta T$ between the consecutive quenching so that $\Delta T/\varepsilon=15 \gg a.$ 
Leaving the details in Appendix \ref{app:PQ-CH-steps}, we found that the evolution is qualitatively similar to Fig.\ref{fig:non-Markov-CH}(a), where the increment in the number of quenched spins corresponds to the {\it reduction} of the irreversibility parameter $a.$ 
This result may be qualitatively understandable because the quenching of 
the spin $s_i$ amounts to the replacement of $\varepsilon_i$ by $\infty,$ or the reduction of $a_i$ to 0 for that spin. 

When the time interval between the consecutive quenching, $\Delta T,$ is not exceedingly larger than either $\varepsilon$ or $\tau,$ the second dimension-free parameter, $\Delta T/\varepsilon,$ comes into play in addition to $a.$ 
Fig.\ref{fig:non-Markov-CH}(b) shows how the final distribution of the total magnetization, $M,$ depends on the values of $\Delta T/\varepsilon,$ where the dimensionless delay $a$ is again fixed at 1.07. 
Note that for $\Delta T/ \varepsilon=0$, the original steady-state of the Choi-Huberman model is recovered as the quenched ensemble after PQ.
With increasing the value of $\Delta T/\varepsilon,$ the free spins have more time to adapt to the quenched part, and the distribution of $M$ undergoes the change which is a qualitatively similar manner to the case of {\it decreasing} the value of $a.$ 

The above results in Figs.\ref{fig:non-Markov-CH}(a) and (b) motivate to study the possible
synthetic effect of $a$ and $\Delta T/\varepsilon,$ or the possible characterization by
$(\Delta T/\varepsilon)/a (=\Delta T/\tau).$ Nevertheless, the comparison on the level of the probability distribution of $M$ is too complicated. We, therefore, characterize each distribution by the second moment $\mathbb{E}[{\null} M^2]$ standardized by its canonical value (i.e., for $a=0$ and arbitrary $\Delta T/\varepsilon$), which we denote by $\mathbb{E}^{(0)}[M^2],$  all knowing that some subtle aspects of the distribution will be lost. For example, the equality, $\mathbb{E}[{\null} M^2] = \mathbb{E}^{(0)}[M^2],$ does not mean that the distribution is identical to the canonical one. Fig.\ref{fig:non-Markov-CH}(c) shows this type of ``projection'' from Fig.\ref{fig:non-Markov-CH}(b), being complemented by more data points. Somewhat surprisingly the ratio $\mathbb{E}[{\null} M^2] / \mathbb{E}^{(0)}[M^2]$ exceeds unity for $\Delta T/\varepsilon\gtsim 8.$ The unimodal-bimodal transition of the distribution takes place where $\mathbb{E}[{\null} M^2] / \mathbb{E}^{(0)}[M^2] =0.9$ approximately (see below).

Fig. \ref{fig:non-Markov-CH}(d) summarizes the contours of $\mathbb{E}[{\null} M^2] / \mathbb{E}^{(0)}[M^2]$ 
on the plane of $a$ and $\Delta T/\varepsilon,$ as the landscape of correlation among quenched spins. Here the total number of spins is $N=4$ because of the limited computing time to ensure good statistics. In fact, $\mathbb{E}[{\null} M^2] / \mathbb{E}^{(0)}[M^2]$ represents rather well the 
characteristics of the probability distribution of $M.$ Especially the unimodal-bimodal transition of the distribution of $M$ is found to occur where $\mathbb{E}[{\null} M^2] / \mathbb{E}^{(0)}[M^2] \simeq 0.9$ (data not shown).
Along the vertical axis with $a=0$, the model is the reversible canonical one, 
therefore, $\mathbb{E}[{\null} M^2] / \mathbb{E}^{(0)}[M^2]=1$ by definition. However, there is another contour of ``canonical'' level. The zone above this contour is ``super-canonical'' realizing $\mathbb{E}[{\null} M^2] / \mathbb{E}^{(0)}[M^2]>1$ although the excess part is very small. This reveals some synergistic effect of the three characteristic time constants, $\varepsilon, \tau$ and $\Delta T.$ The weak undulation of the landscape seen around where the two ``canonical'' contours meet is due to the smallness of the system not being an artifact of statistical error, because (i) the sample size is large enough ($4\times 10^6$) and (ii) the amplitude of undulation decreases when the system size is doubled ($N=8,$ data not shown). Also, the ``super-canonical'' feature is more enhanced, rather than contrary, for the larger system size.

In the parameter region below the second non-vertical ``canonical" contour, 
$\mathbb{E}[{\null} M^2] / \mathbb{E}^{(0)}[M^2]=1,$ the landscape of $\mathbb{E}[{\null} M^2] / \mathbb{E}^{(0)}[M^2]$ is monotone with respect to both $a$ and $\Delta T/\varepsilon.$ 
This suggests that there is a compensating nature of $\Delta T$ for the delay $\tau$. However, near the origin, the perturbation by $a$ is dominant over the influence of $\Delta T/\varepsilon$.

\section{Summary and Remarks}
\label{sec:conlustion}
In the previous works \cite{PQ-KS-BV-pre2018, PQ-CM-KS-2020, PQ-CM-KS-2022} we have studied the ``hidden'' martingale property in the progressive quenching (PQ), that is, the martingale property of the mean of the next quenched spin, associated with the stochastic evolution of the total quenched magnetization.
In the present work, we have demonstrated and numerically verified that the canonical property of the two-story ensemble is behind the martingality and that both the detailed balance and the Markovianity of the stochastic evolution are required for such structure to be maintained.
Under these conditions, the canonicity is conserved even without allowing the unquenched spins to reach a quasi-equilibrium before the subsequent fixation of spins as far as the system starts with a canonical thermal ensemble. (cf. When we go down to a more microscopic scale, the detailed balance may become incompatible with the quenching operation.) When the two-story canonical structure is assured, the hidden martingale holds through the tower-rule applied to the conditional canonical expectations.

We also extended the PQ on the generic transition network having a unique initial steady state. We could characterize sequences of path history (i.e. growing filtration) and constitute the martingale process based on the conditional expectation associated with this filtration.

In the non-Markovian process, even when the system realizes a trajectory-wise detailed balance, the quenching may involve uncontrollable/unobservable modifications in the underlying freedoms that constitute the memory of the observable parts, and such changes can cause the breaking of canonicity of the observable part. 

We also applied the PQ to the system for which the detailed balance is absent even in the unquenched steady ensemble. In the case of PQ on the Choi-Huberman model, the PQ operation can be formulated unambiguously. 
Monitoring through the variance of the total magnetization, we examined the interplay between the intrinsic non-Markovian parameter $\tau$ of the dimension of time and the time interval between the subsequent quenching, $\Delta T$. While the canonical correlation that favored the cooperative fluctuations of spins is attenuated by the non-Markovian delay $\tau$, the operation of quenching reinforces the cooperative fluctuations through $\Delta T.$ 

The last remark is on the similarity and difference between the PQ and some form of ``linear voter models''; see \cite{voter-Liggett} for introduction. 
In a typical example, the binary ($[1,0]$) site (say $x_i$) and its neighbor (say $x_i+n_i$) are chosen at random at each discrete time step and the state of $x$ copies the state of $x+n.$ In that model, $M_t:=\sum_{i=1}^{N_0} x_i(t)/N,$ where $N$ is the system size, is deemed to be either 1 or 0, according to the so-called martingale convergence theorem (see, for example, \cite{klenke2020} Sec. 11.2, "Martingale Convergence Theorems" Example 11.16), while the mean of $M_\infty$ is $M_0/N$ by the martingale property of $M_t.$
If we compare such a model with our PQ of spins, a difference is that 
$M_t$ of the voter model eventually goes only to 1 or 0, unlike our PQ,
while the similarity is that (i) both models have a martingale observable, 
and (ii) the individual realization tends to be polarized due to the interaction with the environment that has a long memory.

\appendix

\section{Details of the two-story ensemble calculations}
\label{app:canonicity}
We show the canonicity of $P_{F, Q}(\mu_T, M_T)$ through a proof by induction.
The key combinatorial identity is the following: 
\begin{align} \label{eq:combidentity}
\binom{T+1}{N_+} \binom{N_0-(T+1)}{n_+}&= \frac{n_+ +1}{N_0-T}\binom{T}{N_+-1}\binom{N_0-T}{n_+ +1} \notag \\
&\!\!\!\!\! +\frac{N_0-T-n_+}{N_0-T}\binom{T}{N_+}\binom{N_0-T}{n_+},
\end{align}
where, as noted in the main text, $N_+=\inv{2}({T+M_{T+1}+1})$  and $n_+ = \inv{2}({N_0-(T+1)+\mu_{T+1}})$ are the number of $(+1)$ spins in the quenched part and unquenched part, respectively.
When $T$ spins have been quenched, we put the hypothesis:
{\small \beqa \label{eq:Pjoint} 
%&& \!\!\!\!\! 
P_{F,Q}(\mu_T,M_T)
%&\equiv & P^{(can)}_{F|Q}(\mu_T|M_T) P_{Q}(M_T)\cr
&=& P^{(can)}_{F,Q}(\mu_T,M_T)
\cr &\equiv & \mathcal{N}_T 
\binom{T}{\frac{T+M_T}{2}} \binom{N_0-T}{\frac{(N_0-T)+\mu_T}{2}}  
e^{\frac{j}{2 N_0}(\mu_T+M_T)^2 },
\cr &&
\eeqa}
where $\mathcal{N}_T$ is the normalization constant such that 
$\sum_{\mu=-(N_0-T)}^{N_0-T\, (mod\, 2)}$
$\sum_{M=-T}^{T\, (mod\, 2)} P_{F,Q}(\mu_T,M_T)=1.$
According to Fig.~\ref{fig:reindexation}, we combine the case of $s_{T+1}=1$ and $s_{T+1}=-1$ as the newly quenched spin with appropriate weight. 
With the above identity (\ref{eq:combidentity}) we can show that the joint probability after the $(T+1)$-th quench, $P_{F,Q}(\mu_{T+1} , M_{T+1})$ is again canonical :
{\small
\beqa
&& P_{F,Q}^{(can)}(\mu_{T+1},T+1 )=
\frac{{N_0-T+\mu_{T+1}+1}}{2(N_0-T)}P_{F,Q}^{(can)}(\mu_{T+1}+1, T), 
\cr &&
P_{F,Q}^{(can)}(\mu_{T+1}, M_{T+1})=
\cr && \qquad
\frac{{N_0-T+\mu_{T+1}+1}}{2(N_0-T)}P_{F,Q}^{(can)}(\mu_{T+1}+1, M_{T+1}-1)
\cr && \qquad
+\,\, \frac{{N_0-T-\mu_{T+1}+1}}{2(N_0-T)}P_{F,Q}^{(can)}(\mu_{T+1}-1 , M_{T+1}+1)
\cr
&& P_{F,Q}^{(can)}(\mu_{T+1},  -(T\! +\! 1))=
\frac{{N_0\! -\! T\! -\! \mu_{T+1}\! +\! 1}}{2(N_0-T)}P_{F,Q}^{(can)}(\mu_{T+1}\!\!
-\!\! 1, -T), 
\cr &&
\eeqa }
where  the first and the last lines apply, respectively for 
$ \mu_{T+1} \ge -(N_0-T)+1$ and $ \mu_{T+1} \le (N_0-T)-1,$
while the middle line applies for $-T\le M_{T+1}\le T.$ 
Because $ P^{}_{F,Q}(\mu_0,M_0=0)= P^{(can)}_{F,Q}(\mu_0,M_0=0)$ by definition,
the proof by induction is completed.
Once we establish $ P^{}_{F,Q}(\mu_T,M_T)= P^{(can)}_{F,Q}(\mu_T,M_T)$ for all $T,$ the marginal $ P^{}_{Q}(M_T)$ is given through $ P^{}_{Q}(M_T)= \sum_{\mu_T} P^{(can)}_{F,Q}(\mu_T,M_T).$

\section{Tower-rule martingale}\label{app:tower}
Let $\{a_1,\ldots, a_N\}:=a_1^N$ be the random variables and 
\beqa \label{eq:mTmtgl}
m_T=m_T(a_{1}^{s}\cup a_{s+1}^{T}) 
&:=& \mathbb{E}[{\null} a_N |a_{1}^{s}\cup a_{s+1}^{T}],
\eeqa
where $a_{1}^{s}\cup a_{s+1}^{T}=a_{1}^{T}.$
Then, by the tower rule of the conditional probabilities, we have
\beqa  
\mathbb{E}[{\null} m_T|a_1^s]&=&
\mathbb{E}[{\null} \,\mathbb{E}[{\null} a_N |a_{1}^{s}\cup a_{s+1}^{T}]\,|a_1^s]
\cr &=& \mathbb{E}[{\null} a_N |a_{1}^{s}]
\cr &=& m_s.
\eeqa
This can be interpreted as the martingality of $m_T$ associated with the process $a_1^T.$ 

\null{
	The tower rule described above supposes that $a_N$ is always the same single random variable. By contrast, in the main text, what we wrote  $s_{N_0}$ in the definition 
	$\meq_T\equiv \mathbb{E}[{\null} s_{N_0}|M_T]$ ($0\le T\le N_0$) is generally the stochastic process, i.e., the sequence of random variables for various ``time'' $T.$
	The $s_{N_0}$ in this definition was implicitly supposed to be measured when the $T$-th spin is quenched. Then, most generally, even though those $\meq_T$ with different $T$ 
	contain the same name, $s_{N_0},$ they concern different $a_N$ in terms of the above formal description.
	Nevertheless, the canonicity of the whole ensemble of both quenched and unquenched spins allows us to lift the distinction of the time at which the expectation of $s_{N_0}$ is taken. This is where the canonicity plays a role.
	We could, therefore, apply the mapping $a_N \mapsto s_{N_0}$ regardless of the time. As a result, the tower rule in this Appendix gives the martingality of $\meq_T.$ 
	In a previous paper \cite{PQ-KS-BV-pre2018}, a derivation of the martingality of $\meq_T$ has been done up to the precision of $\mathcal{O}(N_0)^{-1},$ where they based only on the constrained canonical response $\meq_T.$ The present argument through the tower-rule confirms the link between the canonicity of spin statistics and the martingality of $\meq_T.$
	
	In summary, for the spin model, the replacement by the canonical expectation is indispensable for the martingality, while the availability of homogeneity holds even without resorting to the canonicity.
}
%In fact, $a_T$ could be a random variable of a different nature from the history $a_1^{(T-1)}.$ 

\section{Detailed calculus for 3-Potts model}\label{app:potts}
The picture of the two-story ensemble and the underlying canonical statistics should apply to systems other than the Ising spins on a complete graph. The  $q=3$ Potts model on the complete graph is an example.
Below, we describe the model in some detail.
\subsubsection{Energy and entropy of the $q=3$ Potts model on a complete graph}
The energy of the $q=3$ Potts model on a complete graph reads
\begin{equation*}
	H =-\frac{J_0}{N_0} \sum_{1\le i<j\le N_0} e_i \cdot e_j
	=-\frac{J_0}{2N_0} \left\|{\sum_{1\le i \le N_0} e_i}\right\|^2 +\frac{J_0}{2}, 
\end{equation*}
where $e_i$ is the state of the $i$-th Potts element, etc.
To be concrete, we represent the three states of the Potts' element 
on the plane:  $e^{(1)}=(0,1)^t,$ $e^{(2)}=(\frac{\sqrt{3}}{2},-\inv{2})^t,$ and $e^{(3)}=(-\frac{\sqrt{3}}{2},-\inv{2})^t.$ 
When $T$ of $N_0$ spins have been quenched, their repartition of orientation is denoted by $n^{(i)}$ for the state $e^{(i)}.$ 
We note  $ n^{(1)} +n^{(2)} +n^{(3)} =T$ and $n^{(1)}e^{(1)} +n^{(2)}e^{(2)} +n^{(3)}e^{(3)}=M_T,$ together with  $e^{(1)} +e^{(2)} +e^{(3)}=0.$ The distribution of the $T$ spins can then be characterized by the two parameters:  $\nu^{(1)}=n^{(1)}-n^{(3)}$ and $\nu^{(2)}=n^{(2)}-n^{(3)}.$
By noticing $\nu^{(1)}+\nu^{(2)}=T-3 n^{(3)},$ all $n^{(i)}$ are specified by $\nu^{(1)}$ and $\nu^{(2)}$;
\begin{align*}
	n^{(1)}  &=  \inv{3}({T +2\nu^{(1)}- \nu^{(2)}})\\ 
	n^{(2)} &= \inv{3}({T -\nu^{(1)}+2 \nu^{(2)}}) \\
	n^{(3)} &= \inv{3}({T- \nu^{(1)}- \nu^{(2)}}) 
\end{align*}
With these in mind,
the energy of the whole system reads;
$$
H=-\frac{J_0}{2N_0} \|{M_T +\sum_{T+1\le i \le N_0} e_i}\|^2 +\frac{J_0}{2},
$$
where  $M_T=\nu^{(1)}e^{(1)} +\nu^{(2)}e^{(2)},$ and 
the entropy of quenched part $S(n^{(1)},n^{(2)},n^{(3)})$ (with $\kB\equiv 1$)  reads
$$
e^{S(n^{(1)},n^{(2)},n^{(3)})}= \frac{T!}{n^{(1)}! n^{(2)}! n^{(3)}!},$$
\subsubsection{Consequences of martingality of mean equilibrium spin}
When each node of the complete network has a $q$-state Potts spin with $q>2,$ the homogeneity of the unquenched Potts spins allows, as 
in the Ising ($q=2$) case, the tower-rule-based martingale of the mean equilibrium unquenched spin, which in turn justifies the underlying two-story canonical statistics. Unlike the $q=2$ case \cite{PQ-CM-KS-2022}, the martingale only does not imply the Boltzmann-type weight for the path probability {\it per se.}  Nevertheless, certain constraints are imposed by this martingale property, as will be shown below.

By $\meq_T\equiv \sum_{s_{T+1}} {s_{T+1}} P_T({s_{T}}|M_T),$ the martingale relationship, $\mathbb{E}[{\null} {\meq_{T+1}}|M_T]=\meq_{T}$ reads
{\small
	\beq \label{eq:p-invariance}
	\sum_{s_{T+2}}\sum_{s_{T+1}}({s_{T+2}}-{s_{T+1}})
	P_{T+1}({s_{T+2}}|M_T+{s_{T+1}})  P_T({s_{T+1}}|M_T)=0,
	\eeq}
where we have made use of the identity, 
$\sum_{s_{T+2}}$ $P_{T+1}({s_{T+2}}|M_T+{s_{T+1}}) =1,$ with any
$T$ and $M_T.$ This is valid for any $q\ge 2$ and whatsoever type of symmetric spin-spin interactions such as clock or Potts, etc.
If $q=2,$ this equality immediately goes back to the local invariance 
because the summation contains only those terms with $({s_{T+2}},{s_{T+1}})=(-1,1)$ and $(1,-1).$
For $q=3,$ 
those processes that quench consecutively the two Potts elements of the same state drop out from the summation and 
the above relationship {(\ref{eq:p-invariance})} is reduced to an 2D equality,
that is, starting from a common frozen spin, $M_T=\sum_{i=1}^T s_i,$ there are the two constraints on the local path probabilities of $M_t$: 
\beqa \label{eq:gauge3}
%\exists \phi(M_T)
&& 
P_{T+1,T}(e^{(2)},e^{(1)}|M_T)-P_{T+1,T}(e^{(1)},e^{(2)}|M_T)
\cr & &\,\,\,
= P_{T+1,T}(e^{(3)},e^{(2)}|M_T)-P_{T+1,T}(e^{(2)},e^{(3)}|M_T)
\cr & &\,\,\,
= P_{T+1,T}(e^{(1)},e^{(3)}|M_T)-P_{T+1,T}(e^{(3)},e^{(1)}|M_T).
\eeqa
Each line in the above should represent a common function of $M_T.$

\section{Partial entropy production applied to Landauer's bit memory}
\label{app:Shiraishi}
In this Appendix, we introduce the partial entropy production
\cite{shiraishi-PRE2015} and apply it to Landauer's entropic loss by $\ln 2$ upon the erasure of bit memory. After that, we 
modify this result to show the absence of entropy production when a spin is quenched slowly.

Using the stochastic entropy introduced by Seifert \cite{udo}
the rate of entropy production of a memory bit and the attached heat bath, $ \dot{S}_{tot}=\dot{S}_{Sys}+\dot{S}_{Bath},$ reads 
\beq\label{eq:SdotByP}
\dot{S}_{tot}= \sum_{x} \dot{p_{x}} \ln\frac{q_x}{p_x},
\eeq
where $p_x$ is the probability density at time $t$ while $q_x$ is the steady state probability satisfying the detailed balance (DB),
\beq \label{eq:LDB}
R_{xy}q_y=\tilde{R}_{\tilde{y}\tilde{x}}q_x,
\eeq
with $\tilde{x}$ being the time reversed state of $x$ (when the velocity is included in $x$) and $R_{xy}= W_{x\leftarrow y}$ for $y\neq x,$ and $R_{xx}=-\sum_{y(\neq x)} W_{y\leftarrow x}$ being the minus of the escape rate. By definition, $\sum_x R_{xy}=0.$ By substituting into (\ref{eq:SdotByP}) the master equation,
\beq \label{eq:masterP}
\partial_t p_x = \sum_y R_{xy} p_y
\eeq
and using the DB condition (\ref{eq:LDB}) the partial entropy production $\dot{S}_{x,y}$ specifically associated with the state transition between $x$ and $y$ is given \cite{shiraishi-PRE2015}:
\beq \label{eq:pep}
\dot{S}_{x,y}
:= R_{xy}p_y \ln\frac{{R_{xy}} p_y}{\tilde{R}_{\tilde{y}\tilde{x}} p_x } + \tilde{R}_{\tilde{y}\tilde{x}} p_x - {R}_{xy}p_y.  
\eeq
Here $\dot{S}_{x,y}$ is non-negative because of the generic inequality, $a \ln\frac{a}{b}+b-a\ge 0.$

When the above framework is applied to the double-minimum potential as a model of memory bit, the spatial coordinate $x$ is finely discretized, and the probability density $p_x$ is supposed to be quasi-equilibrium within each valley: $p_x=\theta_t q_x$ for $x< 0$ and $p_x=(2-\theta_t) q_x$ for $x> 0.$ When a memory stocked by this potential is erased through the symmetric lowering of the barrier separating the two valleys, we take $\theta_{t=0}=2$ and $\theta_{t=\infty}=1.$

If we substitute these hypotheses into (\ref{eq:SdotByP}) and integrate over time from 0 to $\infty,$ we already have the expected result, $\ln 2.$ The advantage of the partial entropy production is that (\ref{eq:pep}) allows us to pinpoint where this increment takes place along the potential surface. Except for the vicinity of the barrier top of the potential ($x=0$), the detailed balance (\ref{eq:LDB}) is effectively established, and $\dot{S}_{x,y}$ vanishes, where we supposed the time-reversal symmetry, $\tilde{x}=x$ and $\tilde{R}_{\tilde{y}\tilde{x}}={R}_{{y}{x}}.$ The only transition that can be irreversible is between $x=0-$ and  $x=0^+.$ More concretely,
{\small 
$$
\frac{\dot{S}_{{0^-},{0^+}}}{\kB}\!=\!
R_{{0^-},{0^+}} {\theta_t}q_0  \ln\frac{{\theta_t}}{2-{\theta_t}}
+R_{{0^+},{0^-}}\,p({0^-})-R_{{0^-},{0^+}}\,p({0^+})$$}
{\small $$
\frac{\dot{S}_{{0^+},{0^-}}}{\kB}
\!=\! R_{{0^+},{0^-}}(2-{\theta_t})q_0 
\ln\frac{2-{\theta_t}}{{\theta_t}}
+R_{{0^-},{0^+}}\,p({0^+})-R_{{0^+},{0^-}}\,p({0^-}),$$}
and the sum of these two gives
{\small $$
\frac{\dot{S}_{{0^-},{0^+}}}{\kB}+\frac{\dot{S}_{{0^+},{0^-}}}{\kB}
=[R_{{0^-},{0^+}}{\theta_t}q_0-R_{{0^+},{0^-}}(2-{\theta_t})q_0]
\ln\frac{{\theta_t}}{2-{\theta_t}}.$$}
Here, the quantity in the square bracket on the r.h.s. 
%$[R_{{0^-},{0^+}}q_0\, {\theta_t}-R_{{0^+},{0^-}}q_0\, (2-{\theta_t})]$ 
is the net flow of probability from the left to the right valley. We, therefore, can write
$[R_{{0^-},{0^+}}{\theta_t}q_0-R_{{0^+},{0^-}}(2-{\theta_t})q_0]=\inv{2}|\dot{{\theta_t}}|.$
With this estimation, the time integral of $\dot{S}_{{0^-},{0^+}}+\dot{S}_{{0^+},{0^-}}$  from $\theta_0=2$ to $\theta_\infty=1,$ where $|\dot{{\theta_t}}|=-\dot{{\theta_t}}$, yields finally
{\small $$ \int_{t=0}^\infty dt[
\frac{\dot{S}_{{0^-},{0^+}}}{\kB}+\frac{\dot{S}_{{0^+},{0^-}}}{\kB}]
=\int_0^\infty (-\inv{2}\dot{{\theta_t}})\ln\frac{{\theta_t}}{2-{\theta_t}}dt =\ln 2.$$}
Therefore, only the partial entropy production at the top of the potential barrier is responsible for all the entropy loss.

When we retrace the above reasoning in our Progressive Quenching (PQ) case, the total entropy production should also be concentrated at the barrier top, $x=0^\pm,$ where the barrier height is raised to well above $\kT.$ If the DB is maintained during the operation of PQ, the current $[R_{{0^-},{0^+}}{\theta_t}q_0-R_{{0^+},{0^-}}(2-{\theta_t})q_0]$ vanishes, unlike the case of Landauer, and the production, $\dot{S}_{{0^-},{0^+}}+\dot{S}_{{0^+},{0^-}},$   does also. In conclusion, when the potential barrier is raised slowly enough on the microscopic scale, the operation of PQ is reversible, and, as a result, the two-story canonical distribution is maintained.

\section{Discrete-time version of Section \ref{subsec:Markov_DB}} \label{an:Discrete_time_PQ_on_TN}
We consider a \textit{discrete-time} and discrete-state Markov process characterized by the transition probabilities $K(\alpha_i {\!\to} \alpha_j).$ We suppose that this process is stationary, and we denote by $P^\mathrm{{st}}(\alpha)$ the stationary probability.
We will show that if the detailed balance (DB) holds in the stationary ensemble, the Progressive Quenching (PQ) allows $P^\mathrm{{st}}(\alpha_i)$ to remain the stationary distribution.

When a group of degrees of freedom, say $\{a_i\}^Q,$ are quenched, certain transitions
that involve the change of this variable are prohibited. We introduce 
$\delta^Q(\alpha,\alpha')$ so that $\delta^Q(\alpha,\alpha')=1$ [$0$] if the transitions
$\alpha{\!\to}\alpha'$ and $\alpha'{\!\to}\alpha$ are allowed [prohibited], respectively.
Then under the condition of quenched variables, $\{a_i\}^Q,$ the off-diagonal transition
probabilities, that we denote by $K^Q(\alpha_i {\!\to} \alpha_j)$ ($i\neq j$), should read
$$
K^Q(\alpha_i {\!\to} \alpha_j)=\delta^Q(\alpha,\alpha') K(\alpha_i {\!\to} \alpha_j)
$$
To maintain the normalization condition of the probability, the diagonal element
of the transition probability should also be compensated:
\beqa
&&K^Q(\alpha_i {\!\to} \alpha_i)
\cr &&=
1-\sum_j^{(j\neq i)} K^Q(\alpha_i {\!\to} \alpha_j)
\cr &&=K(\alpha_i {\!\to} \alpha_i)
+\sum_j^{(j\neq i)} (1-\delta^Q(\alpha_i,\alpha_j))K(\alpha_i {\!\to} \alpha_j)
\eeqa
We now ask whether the stationary distribution of the original transition network (TN),
$P^\mathrm{{st}}(\alpha_i),$ remains so for the quenched TN if the former TN satisfies the DB. We rewrite the stationarity condition of $P^\mathrm{{st}}(\alpha_i)$ :
\beqa \label{eq:PofQ}
&& 
P^\mathrm{{st}}(\alpha_i)
%\cr &&
= \sum_j  P^\mathrm{{st}}(\alpha_j)K(\alpha_j {\!\to} \alpha_i).
\cr &&=
\sum_{j(\neq i)}  P^\mathrm{{st}}(\alpha_j)
\inRbracket{\delta^Q(\alpha_i,\alpha_j)+[1-\delta^Q(\alpha_i,\alpha_j)]}
K(\alpha_j {\!\to} \alpha_i)
\cr && \,\,\,+P^\mathrm{{st}}(\alpha_i)K(\alpha_i {\!\to} \alpha_i)
\cr &&= \sum_{j(\neq i)}  P^\mathrm{{st}}(\alpha_j) K^Q(\alpha_j {\!\to} \alpha_i)
\cr &&\,\,\,+
\sum_{j(\neq i)}[1-\delta^Q(\alpha_i,\alpha_j)]
P^\mathrm{{st}}(\alpha_j)K(\alpha_j {\!\to} \alpha_i)
+
P^\mathrm{{st}}(\alpha_i)K(\alpha_i {\!\to} \alpha_i)
\cr &&= \sum_{j(\neq i)}  P^\mathrm{{st}}(\alpha_j) K^Q(\alpha_j {\!\to} \alpha_i)
\cr &&\,\,\,+
P^\mathrm{{st}}(\alpha_i)
\sum_{j(\neq i)}[1-\delta^Q(\alpha_i,\alpha_j)] K(\alpha_i {\!\to} \alpha_j)
+ P^\mathrm{{st}}(\alpha_i)K(\alpha_i {\!\to} \alpha_i)
\cr &&= \sum_{j(\neq i)}  P^\mathrm{{st}}(\alpha_j) K^Q(\alpha_j {\!\to} \alpha_i)
+P^\mathrm{{st}}(\alpha_i)K^Q(\alpha_i {\!\to} \alpha_i)
\cr &&= \sum_{j}  P^\mathrm{{st}}(\alpha_j) K^Q(\alpha_j {\!\to} \alpha_i),
\eeqa
where the fourth equality is due to the DB condition of the original TN,
$$
P^\mathrm{{st}}(\alpha_j)K(\alpha_j {\!\to} \alpha_i)
=P^\mathrm{{st}}(\alpha_i)K(\alpha_i {\!\to} \alpha_j).
$$
Eq.(\ref{eq:PofQ}) means that $P^\mathrm{{st}}(\alpha_i)$ is a stationary distribution
of the quenched system, though it may not be the unique one.

\section{Hidden-Spin Model}
This Appendix gives some details of what we recall in Sec.\ref{sub2sec:recall} of the main text.
\subsection{Non-Markovianity of the Hidden-Spin model}\label{subsec:Non_Markov_of_hidden_spins}
Let the system have the energy function, 
$$ H= -K (s_1+s_2)\sigma,$$
where $K>0$ and the variables, $s_1,s_2$ and $\sigma,$ are Ising spins. 
The ``hidden'' spin $\sigma$ mediates the interaction between the  ``observable'' spins, $s_1$ and $s_2.$ 
We shall use the unit such that the inverse temperature is $\beta=1.$ We assume a Markovian evolution of this system but observe only $s_1$ and $s_2.$ 
We introduce a short time step, $dt,$ and focus on the three consecutive instants, $\{t_{k-1},t_{k},t_{k+1}\}=\{(k-1)\, dt,k\, dt,(k+1)\, dt\}.$ We also introduce the notations,
$\alpha_{k}= (s_1(t_k), s_2(t_k))$ and $\sigma_{k}=\sigma(t_k)$.
Supposing $dt\ll \epsilon,$ we will ignore the errors of $\mathcal{O}((dt)^2).$ 
Our main concern is the history-conditioned probability, 
$P(\alpha_{{k}+1}|\alpha_{k},\alpha_{{k}-1}),$ and we claim the general inequality, 
$P(\alpha_{{k}+1}|\alpha_{k},\alpha_{{k}-1})\neq P(\alpha_{{k}+1}|\alpha_{k},\alpha^\prime_{{k}-1})$  for $\alpha_{{k}-1}\neq \alpha^\prime_{{k}-1}.$
The Markovian transition probability in terms of the micro-state 
$\left(  {\pair{\alpha_{{k}}}{\sigma_{{k}}}}\right)$ is written as 
$P\left.\left(\pair{\alpha_{{k}+1}}{\sigma_{{k}+1}}\right| \pair{\alpha_{{k}}}{\sigma_{{k}}}\right).$ 
Using this, the conditional probability $P(\alpha_{{k}+1}|\alpha_{k},\alpha_{{k}-1})$
can be expressed as
{\small
	\beqa 
	&& P(\alpha_{{k}+1}|\alpha_{k},\alpha_{{k}-1})
	\cr &&
	= \frac{P(\alpha_{{k}+1},\alpha_{k},\alpha_{{k}-1})}{P(\alpha_{k},\alpha_{{k}-1})}
	\cr &&
	=\frac{	\sum_{\sigma_{{k}+1},\sigma_{{k}},\sigma_{{k}-1}}
		P\left.\left(\pair{\alpha_{{k}+1}}{\sigma_{{k}+1}}\right|
		\pair{\alpha_{{k}}}{\sigma_{{k}}}\right)
		P\left.\left(\pair{\alpha_{{k}}}{\sigma_{{k}}}\right| 
		\pair{\alpha_{{k}-1}}{\sigma_{{k}-1}}\right)
		P\left(\pair{\alpha_{{k}-1}}{\sigma_{{k}-1}}\right)
	}{
		\sum_{\sigma'_{{k}},\sigma'_{{k}-1}} 
		P\left.\left(\pair{\alpha_{{k}}}{\sigma'_{{k}}}\right| \pair{\alpha_{{k}-1}}{\sigma'_{{k}-1}}\right)P\left(\pair{\alpha_{{k}-1}}{\sigma'_{{k}-1}}\right)
	}.
	\cr &&
	\eeqa}
As for the probability $P\left(\pair{\alpha_{{k}-1}}{\sigma_{{k}-1}}\right)$ 
we assume the canonical weight, $\exp\left(-H\left(\pair{\alpha_{{k}-1}}{\sigma_{{k}-1}}\right)\right)/Z$ with the partition function, $Z=4+2 e^{2K}+2 e^{-2K}.$ 
In the conditional probability 
$P\left.\left(\pair{\alpha_{{k}+1}}{\sigma_{{k}+1}}\right| \pair{\alpha_{{k}}}{\sigma_{{k}}}\right)$ 
we ignore the flipping of more than one spin because such an event weights $\mathcal{O}((dt)^2).$ 
As for the single spin flip, we use the formalism of Bergmann-Lebowitz \cite{Lebowitz55}: 
The transition rate $W_{b\leftarrow a}$ from the micro-state $a$ to $b$ 
for $P(b|a)=W_{b\leftarrow a} dt$ takes the form,
$W_{b\leftarrow a}=\nu_0 e^{-(\Delta_{[a,b]} -F_a)},$ which assures the (microscopic) Markovian DB.
%whose landscape picture is clear. 
For those transition flipping $\sigma$ we assign $\Delta_{[a,b]}=\delta$ while for those keeping $\sigma$ fixed we assign $\Delta_{[a,b]}=\Delta.$ The energy value $F_a$ takes among 
$\{-2K,0,2K\}.$ Some symbolic calculus tells that,
%The Mathematica\footnote{cf. {\sf non-Markov.nb}} tells about $P(\alpha_{{k}+1}|\alpha_{k},\alpha_{{k}-1})$ as follows:  While 
while the transitions from the anti-parallel pair do not reflect the past further than $\mathcal{O}(dt)$\footnote{The notation (+-) means $(s_1,s_2)=(1,-1),$, etc.};
\begin{align}
	P((++)|(+-),(\bm{++}))&= P((--)|(+-),(\bm{++})) \notag 
	\\ &=(\nu_0 dt) e^{-\Delta} \notag \\
	P((++)|(+-),(\bm{+-}))&= P((--)|(+-),(\bm{+-})) \notag 
	\\ &= (\nu_0 dt) e^{-\Delta},
\end{align}
the transition from the parallel pair depends on the further past;\footnote{Recall that at most only one spin can flip during $dt.$ Therefore, $\alpha_{{k}+1}\neq \alpha_{k}$ implies $\sigma_{{k}+1}=\sigma_{k}.$}
\begin{align}
	\label{eq:nonMarkovProb3}
	P((-+)|(++),(\bm{-+}))&=P((+-)|(++),(\bm{-+})) \notag
	\\ &=(\nu_0 dt) e^{-\Delta} \cosh (2K) \notag \\
	P((-+)|(++),(\bm{++}))&= P((+-)|(++),(\bm{++})) \notag
	\\ &= (\nu_0 dt) e^{-\Delta} \sech (2K).
\end{align}
Thus, we claim the general inequality;
$P(\alpha_{{k}+1}|\alpha_{k},\alpha_{{k}-1})\neq P(\alpha_{{k}+1}|\alpha_{k},\alpha^\prime_{{k}-1})$  for $\alpha_{{k}-1}\neq \alpha^\prime_{{k}-1}.$
Intuitively, Eq.(\ref{eq:nonMarkovProb3}) means that if the state $(++)$ is  realized only during $[(k-1)dt,k dt],$ the transition to $(-+)$ or to $(+-)$
is enhanced by the factor $\coth(2K) \,(>1)$ as compared with     
the case in which the state $(++)$ has been maintained before $(k-1)dt.$ 

\subsection{Trajectory-wise detailed balance}\label{an:Trajectory_DB}
Suppose a system undergoes a Markovian stochastic process and satisfies the detailed balance (DB).
We denote by $\omega_k=(\pair{\alpha_k}{\sigma_k})$ the state of the system at time $t_k,$ where $\alpha$ and $\sigma$ stand for the visible and hidden variables, respectively.
For simplicity the time-reversed state of $\omega$ is assumed to be $\omega.$
As in Sec.~\ref{subsec:Non_Markov_of_hidden_spins} we use the discretization of time with small interval $dt.$ 
Then the ``instantaneous'' DB condition for $\omega,$ reads, 
\eqn{\label{eq:DBinst}
{P}(\omega_{k-1}|\omega_{k})P^{eq}_{\omega_k}= {P}(\omega_{k}|\omega_{k-1})P^{eq}_{\omega_{k-1}}.
}
In using repeatedly this relation, we have the trajectory-wise DB for the variable $\omega$
starting from the canonical state, $P^{eq}:$
\begin{align} \label{eq:pathDBssigma}
\mathbb{P}(\{\omega_k\}_{k=0}^\mathcal{K})
&={P}(\omega_{\mathcal{K}}|\omega_{\mathcal{K-1}})\cdots 
{P}(\omega_2|\omega_1) {P}(\omega_1|\omega_0) 
P^{eq}_{\omega_0\mathcal{}} \notag\\ 
&= P^{eq}_{\omega_\mathcal{K}} {P}(\omega_{\mathcal{K}-1}|\omega_{\mathcal{K}})\cdots  {P}(\omega_1|\omega_2) {P}(\omega_0|\omega_1) \notag\\
&={P}(\omega_0|\omega_1) {P}(\omega_1|\omega_2) \cdots {P}(\omega_{\mathcal{K}-1}|\omega_{\mathcal{K}})P^{eq}_{\omega_\mathcal{K}} \notag\\ 
&\equiv  \mathbb{P} (\{\omega_k\}\!{^*}_{k=0}^\mathcal{K}).
\end{align}
We then focus only on the history of the visible observables, $\{\alpha_k\}_{k=0}^\mathcal{K}.$ For that purpose we integrate out the hidden part, $\{\sigma_k\}_{k=0}^\mathcal{K}:$ 
$$\mathbb{P}(\{\alpha_k\}_{k=0}^\mathcal{K}) = \sum_{\{\sigma_m\}_{m=0}^\mathcal{K}} \mathbb{P}( \{\omega_k\}_{k=0}^\mathcal{K}\,), 
$$ 
where the sum is taken under the fixed $\{\alpha_k\}_{k=0}^\mathcal{K}.$
Applying (\ref{eq:pathDBssigma}) to each term on the r.h.s. above, we have
\begin{align}
l.h.s. &=
\sum_{\{\sigma_m\}{^*}_{m=0}^\mathcal{K}} \mathbb{P}( \{\omega_k\}{^*}_{k=0}^\mathcal{K}\,) \\
&= \mathbb{P}(\{\alpha_k\}{^*}_{k=0}^\mathcal{K}).
\end{align}
Thus, we have the trajectory-wise DB relation.
\begin{equation} \label{eq:annex_DB_for_obs_spins}
\mathbb{P}(\{\alpha_k\}{}_{k=0}^\mathcal{K})=\mathbb{P}(\{\alpha_k\}{^*}_{k=0}^\mathcal{K}).
\end{equation}
The relation like (\ref{eq:DBinst}) does not hold any more because 
$\mathbb{P}(\alpha_{k+1},\alpha_{k})={P}(\alpha_{k+1}|\alpha_{k})P^{eq}_{\alpha_k}$ contains behind many trajectories of $\omega.$
\section{Absence of Detailed Balance in the Choi-Huberman model}\label{an:22CH}
\begin{figure}[t!!]
\centering\includegraphics[width = 0.85\linewidth]{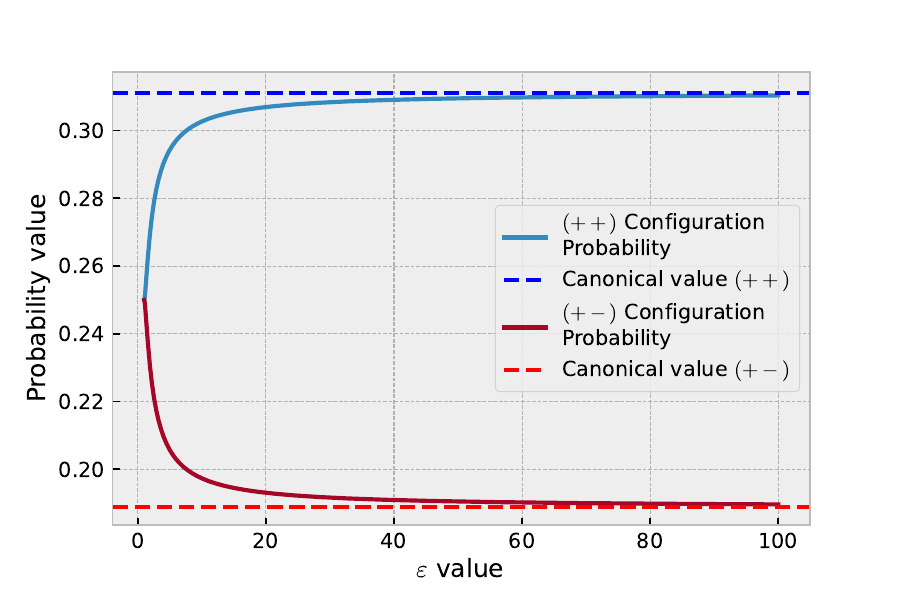}
\caption{Plot of the steady-state probabilities (solid curves):
	$P^{{\rm st}}({\tiny\begin{pmatrix} +1 \\ +1 \end{pmatrix}}) =
	P^{{\rm st}}({\tiny\begin{pmatrix} -1 \\ -1 \end{pmatrix}})$ in blue, and $P^{{\rm st}}({\tiny\begin{pmatrix} +1 \\ -1 \end{pmatrix} })=P^{{\rm st}} ({\tiny \begin{pmatrix} -1 \\ +1 \end{pmatrix} })$ in red, as functions of $\varepsilon (\ge 1).$     
	$\eta=\tanh(\beta j)$ has been chosen at $\beta j=1/4.$
	Data were drawn using the formula (\ref{eq:PstCH}).}
\label{fig:CH22_epsilon}
\end{figure}

This Appendix focuses on verifying the non-canonical nature of the steady state of the Choi-Huberman model, the Ising spin system with delayed interaction. The approach is to take up a simplified and discrete-time version of the Choi- Huberman (CH) model applied to the two Ising spins and to show that its steady state depends on the kinetic parameters, which is not the case for the canonical ensemble.

The time is discretized with the unit being unity, and we set the delay of the interaction $\tau$ to be equal to this unit, i.e. $\tau=1.$ 
We denote by $\vec{s}(t):=\begin{pmatrix} {s}_{0}(t) \\ {s}_{1}(t) \end{pmatrix}$ the polarization of the two spins at time $t.$ Following the CH model, the probability that ${s}_{0}(t)$ [${s}_{1}(t)$] at $t$ are flipped at the time $(t+1)$ depends on the spin state of their (exclusive) neighbors but 
at the time $(t-1)$, that is, on ${s}_{1}(t-1)$ [${s}_{0}(t-1)$]. In the spirit of the (discrete-time) Glauber model, we adopt the probabilities,
\begin{align}
P\left[ {s}_{0}(t+1) = -{s}_{0}(t)| \vec{s}(t),\vec{s}(t-1) \right]  &= \frac{1}{2\varepsilon}(1-{s}_{0}(t){s}_{1}(t-1)\eta)\\
P\left[ {s}_{1}(t+1) = -{s}_{1}(t)|\vec{s}(t),\vec{s}(t-1) \right]  &= \frac{1}{2\varepsilon}(1-{s}_{1}(t){s}_{0}(t-1)\eta),
\end{align}
with $\eta = \tanh(\beta j/2),$ where $j/2$ is the coupling constant for this two spin system,
similarly to the Glauber mode. (The factor 1/2 is merely from the convention on the complete graph applied to $N=2$ spins.)
For the sake of simplicity, we assume that the flip of the two spins takes place
independently for one from the other. Then the probability of $\vec{s}(t+1)$ reads, 
\beqa\label{eq:CH2spin}
&& P(\vec{s}(t+1)|\vec{s}(t),\vec{s}(t-1)) 
\cr
&& = \frac{1}{4}  \left[  1+{s}_{0}(t+1){s}_{0}(t)\left( 1 - \frac{1- {s}_{0}(t){s}_{1}(t-1)\eta}{\varepsilon}\right)\right] 
\cr && 
\times
\left[  1+{s}_{1}(t+1){s}_{1}(t)\left( 1 - \frac{1- {s}_{1}(t){s}_{0}(t-1)\eta}{\varepsilon}\right)\right],
\label{eq:an_probaCH22}
\eeqa
where, for the discrete-time version, Glauber's elementary time scale, $\varepsilon,$ should be no less than unity.
To handle the above non-Markovian transition probabilities, 
we follow the usual technique of Markovianizing the original description by extending the state so that the state involves more than one moment. Here, we choose as the extended state the $2\times 2$ matrix, 
$(\vec{s}(t+1),\vec{s}(t)):=\begin{pmatrix}
{s}_{0}(t+1) & {s}_{0}(t) \\ {s}_{1}(t+1) & {s}_{0}(t)
\end{pmatrix}.$ 
Through such redefinition of the state, 
the transition from $(\vec{s}(t),\vec{s}(t-1))$ to $(\vec{s}(t+1),\vec{s}(t))$ is Markovian.  
(cf. The redundancy of the description due to the reappearance of $\vec{s}(t)$ does not
harm the procedure.) Formally, this Markov chain should be written as a $16\time 16$ matrix, 
and the question of the steady state probability is reduced to the search of the eigenvector of this matrix with the unitary eigenvalue. After some symbolic calculus,
the steady-state probability $P^{\rm st}(\vec{s})$ is found to be 
\beqa \label{eq:PstCH}
P^{{\rm st}}({\tiny\begin{pmatrix} +1 \\ +1 \end{pmatrix}}) 
&=&  
\frac{(1+\eta) (\varepsilon-\eta)^2 (2 \varepsilon+\eta-1)}{8 \varepsilon^3-4 \varepsilon^2 \left(3 \eta^2+1\right)+8 \varepsilon \eta^2+4 \eta^2 \left(\eta^2-1\right)}
\cr &=& P^{{\rm st}}({\tiny\begin{pmatrix} -1 \\ -1 \end{pmatrix}})
\cr 
P^{{\rm st}}({\tiny\begin{pmatrix} +1 \\ -1 \end{pmatrix} }) &=& 
\frac{(1-\eta) (\varepsilon+\eta)^2 (2 \varepsilon-\eta-1)}{8 \varepsilon^3-4 \varepsilon^2 \left(3 \eta^2+1\right)+8 \varepsilon \eta^2+4 \eta^2 \left(\eta^2-1\right)}
\cr  &=& P^{{\rm st}} ({\tiny \begin{pmatrix} -1 \\ +1 \end{pmatrix} }).
\eeqa
Evidently, the steady state depends on the kinetic parameter $\varepsilon,$ as a sign of 
non-canonical ensemble.
Figure \ref{fig:CH22_epsilon} shows the above probabilities as function of $\varepsilon (\ge 1).$ 
In the limit, $\varepsilon \to +\infty,$  the system behaves canonically (the top and bottom dashed lines), whereas in the limit, $\varepsilon \to 1^+,$ all the probabilities become  1/4.
The theoretical formula (\ref{eq:PstCH}) is also in excellent agreement with numerical simulations.

\section{Evolution of spin correlations along PQ in the two-story ensemble}
\label{app:PQ-CH-steps}
\begin{figure}[h!]
	\centering
	\includegraphics[width = 0.85\linewidth]{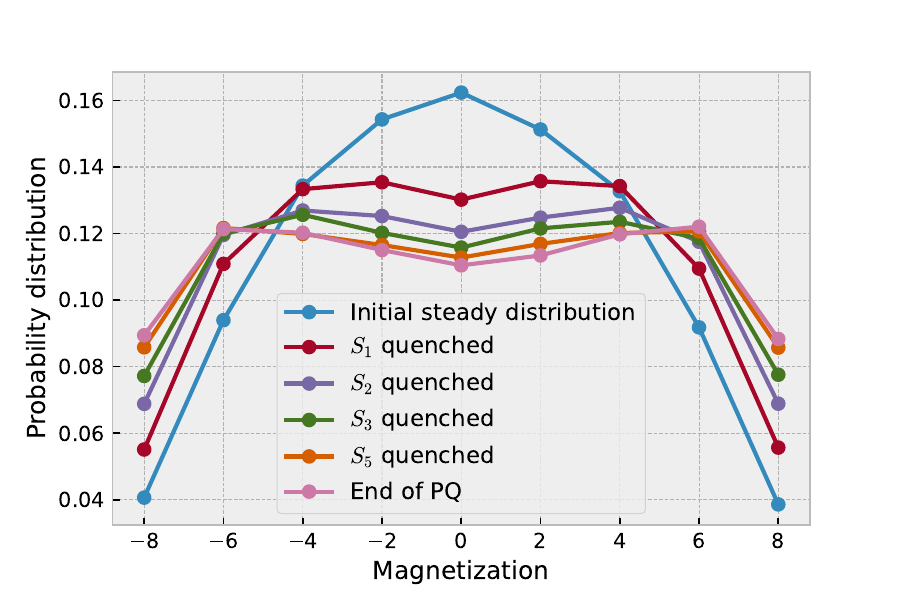}
	\caption{ The distribution of magnetization at different stages of PQ. The non-dimensionalized delay $a$ and the non-dimensionalized time interval between consecutive quenches, $\Delta T/\varepsilon,$ are fixed at $(a, \Delta T/\varepsilon)=(1.07,15)$. }
	\label{fig:PQ_Mag_evolution}
\end{figure}

Fig.\ref{fig:PQ_Mag_evolution} shows how the distribution of the total magnetization evolves with the stages of PQ. 
Here, it is understood that, when a part of all of the spins are quenched, the total magnetization $M$ is calculated using the two-story ensemble, including both quenched spins and thermally fluctuating ones (see Sec.\ref{subsec:two-story}). The kinetic parameters are fixed at $(a,\Delta T/\varepsilon)=(1.07,15).$ As compared with the CH model without quenching (the unimodal blue points and links), we observe that the progress of quenching enhances the correlation among the spins, as is the case with a decreased delay parameter $a$ observed in Figs.\ref{fig:non-Markov-CH}(a). 

\section{An extended PQ on the transition network and martingale}\label{app:TNmartingale}
\begin{figure}[h!!]
	\includegraphics[width = \linewidth]{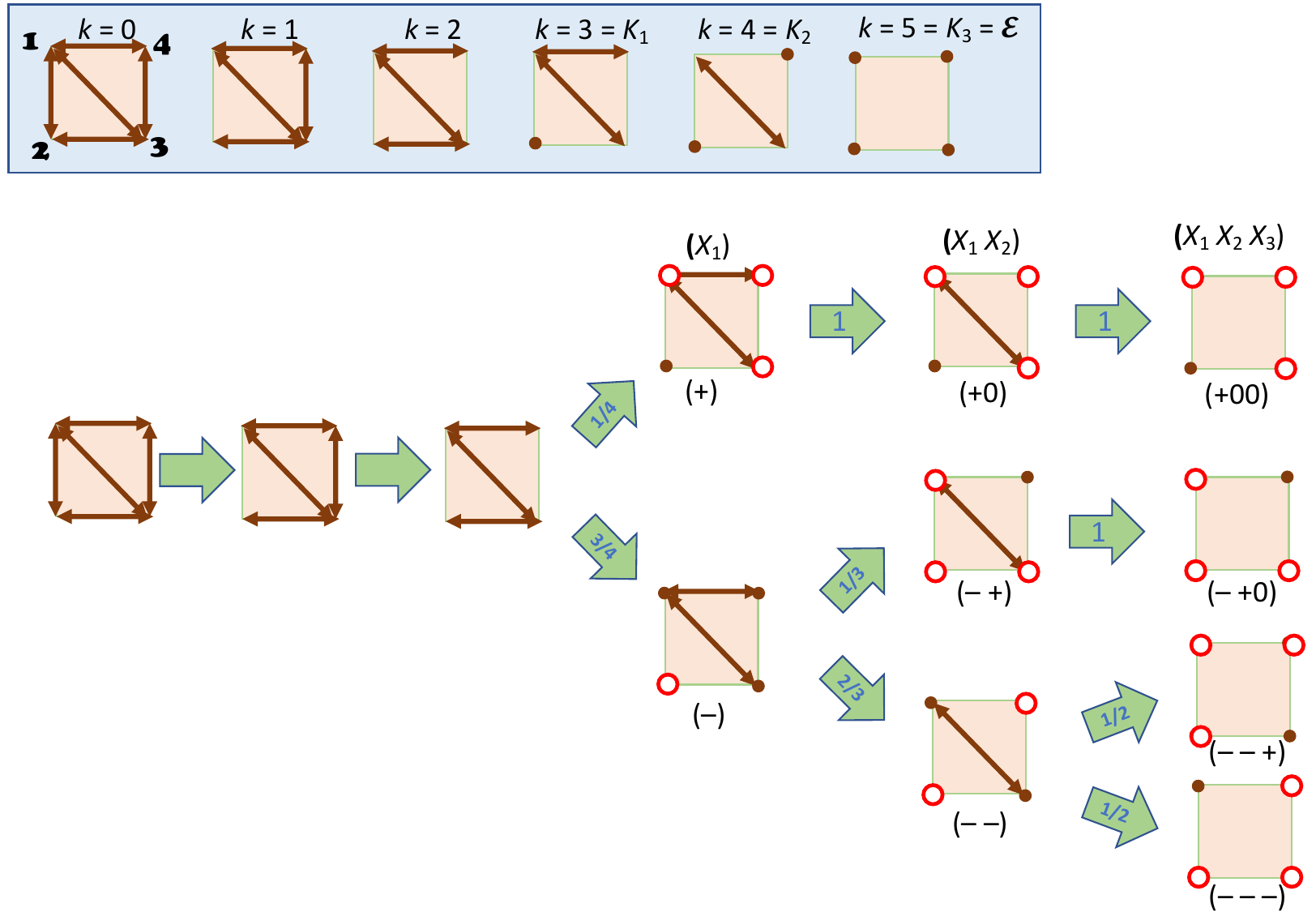}
	\caption{A simple example of Progressive Quenching (PQ) on the Markovian transition network (TN) with non-canonical steady state. Here $\# (\mathrm{edge})=\mathcal{E}=5$ and $\# (\mathrm{vertex})=V=4.$ The initial TN is shown in leftmost within the top box. The bidirectional transition rates on each edge have been chosen uniformly randomly from the interval $[0,1].$ The PQ of the TN is shown in this box from left to right, {where} $k$ counts the ``time'' at which the $k$-th bidirectional edge is removed, while $k=K_n$ indicates the occurrence of the $n$-th ergodicity breaking of TN. After each quench, we give enough (real) time so that the distribution is stationary within each ergodic island. In the main figure, we show the branching of the path ensemble (i.e. growing filtration) by thick arrows in which the partition probabilities are also noted. (cf. This is a stochastic process while the sequence shown in the top box is deterministic.) 	The filtrations are indicated by the process $\{X_1,\ldots, X_n\}$ ($n\ge 1$), where we used the notation $X_n=\pm$ instead of $(\pm 1)$ for the sake of visibility. Those thick arrows in orange color mean the change of probability flow on the TN. Those pathways with zero transition rates have been omitted. The red open circle indicates the absence of a visit. The martingale process $Y_n$ with respect to $\{X_1,\ldots, X_n\}$ can be constructed in the form of $Y_n=\mathbb{E}[{\null} z|\{X_1,\ldots, X_n\}]$ with $z$ being a stochastic process adapted to $\{X_1,\ldots, X_n\}.$
	}
	\label{fig:PQ-TN-mtgl}
\end{figure}

\null{ 
	We consider, thanks to a creative question of an anonymous reviewer, a generalized PQ that applies to any finite Markovian transition network (TN). The edges of the network are supposed to bear bidirectional transition rates such that the TN initially has a unique steady state. 
	We progressively remove bidirectional transition edges ($\mathcal{E}$ in total) in the TN following a given protocol until all the individual state ($V$ in total) becomes isolated. We then ask whether there is a concomitant martingale process. 
	When we compare the martingality by the tower property of the conditional expectation, $\meq_{T, M_T},$ (see Appendix \ref{app:tower}) with the martingality of the path probability ratio, which leads to the integral fluctuation theorems - see \cite{martingales_review2022} \S~2.1.3, around Eq.(2.15) -  we notice that the conditional probabilities and the progressive specifications of histories $\{X_1,\ldots, X_n\},$ or the {\it growing filtration} for mathematicians, are the underlying common basis.
	Therefore, it will be natural to keep these notions in mind when extending the PQ martingale to the transition network. Below is a proposition along this line.
	
	We introduce the ``microscopic time'' $k(= 1,\ldots, \mathcal{E})$ at which the $k$-th bidirectional edge is removed (``quenched'') from the TN. We allow for enough (physical) time between consecutive quenches so that the probability flow becomes stationary. This protocol is fixed.
	We denote by $\{\omega\}$ the whole histories of transitions on the TN, each of which obeys the kinetic constraints imposed by the PQ.
	
	We then introduce $K_1, K_2,\ldots$ such that $k=K_n$ is the $n$-th ``microscopic time stamp" at which a cluster of TN is subdivided into two separate clusters. There are $(n+1)$ isolated clusters after the time $k=K_n.$ 
	The index $n,$ therefore, plays a role of ``macroscopic time'' for the progress of quenching. 
	Even without detailed balance, the ``neutrality'' of the PQ is assured since, upon the non-ergodic transition the probability-flow through the so-called {\it bridge bond} to be removed should have already been zero in the steady state.
	We also notice $\max{\![\,n\,]}=V-1$ since the number of clusters 
	will not exceed the number of states, $V.$ Besides, the last division must eliminate the last bridge bond. Therefore,
	\beq\label{eq:KVE}
	K_{V-1}={\mathcal{E}}.
	\eeq
	For a particular history $\omega,$ the $n$-th sub-division may or may not happen in the cluster where $\omega$ finds itself at that time $K_n.$ When it does, this history $\omega$ should continue in one of the two newly sub-divided clusters, to which we assign $X_n(\omega)=\pm 1,$ while when it does not, we assign $X_n(\omega)=0.$ 
	In this manner, we can associate a sequence $\{X_1,\ldots, X_{V-1}\}$ ($X_n\in \{-1,0,1\}$) 
	for each history $\omega.$\footnote{By $(X_1,\ldots, X_n)$ we know in which ergodic island the actual system belongs to after $n$-th ergodicity breaking. }
	For $n=1$ we note that $X_1\neq 0$ because any history $\omega$ must experience the division of the whole TN at $k=K_1.$ In fact $\{X_1,\ldots, X_{V-1}\}$ can completely specify the itinerary of $\omega$ on the cluster level, i.e. 
	$\{X_1,\ldots, X_n\}$ ($1\le n\le V-1$) constitutes a growing filtration. In Fig.\ref{fig:PQ-TN-mtgl}, we show a simple example of PQ on Markovian TN.\\
	\mbox{}\\
	
	These having been prepared, the conditional expectation, 
	\begin{equation}
		Y_n\equiv \mathbb{E}[{\null} z|X_1,\ldots,X_n]
	\end{equation}
	is martingale with respect to the process  $\{X_1,\ldots,X_n\},$ where $z$ is a random variable as function of $\omega.$ 
	
	As a trivial example, $z$ can be the binary indicator 
	that tells whether the final state of $\omega$ at $k=\mathcal{E}$ belongs to a prefixed subset of states, $A:$ 
	\begin{equation}
		z={\bf 1}_{\omega_{k=\mathcal{E}}\in A}.
	\end{equation}
	Then $Y_n$ is the conditional probability, 
	\begin{equation}\label{eq:PofAasYn}
		Y_n =P({\omega_{k=\mathcal{E}}\in A}|X_1,\ldots,X_n).
	\end{equation}
	If we regard $Y_n$ as function of the subset $A$ and further chose as $A$ the individual state $\{a\}$ on the TN, then $p_n(a)\!:=Y_n$ gives the {\it probability distribution} of final destinations given the early partial itinerary, $(X_1,\ldots, X_n).$
	The martingality, 
	$$\mathbb{E}[{\null} p_r (a)|X_1,\ldots,X_n]=p_n(a),\quad (r\ge n).$$
	is then nothing but the tower-rule structure of the {\it conditional probability.}
}
 % canonical-quench.tex

\null{ In the Fig.\ref{fig:PQ-TN-mtgl} where $(\mathcal{E}, V)=(5,4),$ we took $z={\bf 1}_{\omega_{5}={\bf 1}},$ where $Y_n$ means the conditional probability for a process to end at the state {\bf 1} subjected under the conditions $\{X_1,\ldots, X_n\}$ that specify those non-ergodicity transitions undergone up to the $n$-th one.\footnote{For a general $z$ the process $Y_n$ is non-Markovian because, to specify an ergodic island from the data
		$\{X_1,\ldots,X_n\},$ we have to find the $m$ such that $|X_m|\neq 0$ and 
		$X_{m+1}=\cdots = X_n=0.$}
	We can verify by hand, for example,
	$\mathbb{E}[{\null} Y_3|X_1]=Y_1(X_1)$ with $X_1=(-1).$ (The case for $X_1=+1$ is trivial.) We evaluate
	$\mathbb{E}[{\null}  \mathbb{E}[{\null} {\bf 1}_{\omega_{5}={\bf 1}}|X_1=-1,X_2,X_3]\,|X_1=-1]$ by taking the expectations for $(X_2,X_3)=(-1,1,0),(-1,-1,1)$ and $(-1,-1,-1).$ 
	We end up with comparing $\mathbb{E}[{\null} Y_3|X_1=-1]=0.712\times 0.814 (\simeq 0.58)$ with $Y_1(X_1=-1)=0.452/0.775 (\simeq 0.58).$)
}

\clp
\bibliographystyle{apsrev4-2}

\end{document}